\documentclass{aa}  
\usepackage{graphicx}
\usepackage{txfonts}
\usepackage[]{hyperref}
\begin{document}

\title{SUPPNet: Neural network for stellar spectrum normalisation}

\authorrunning{R\'o\.za\'nski et al.}

   \author{T. R\'o\.za\'nski,\inst{1}
           E. Niemczura,\inst{1}
           J. Lemiesz,\inst{2}
           N. Posi{\l}ek,\inst{1} 
          \and
           P. R\'o\.za\'nski\inst{3}
          }

   \institute{Astronomical Institute, University of Wroc\l aw, 
             Kopernika 11, 51-622 Wroc\l aw, Poland\\
              \email{tomasz.rozanski@uwr.edu.pl}
         \and
             Department of Computer Science, Faculty of Fundamental Problems of Technology, Wroc{\l}aw University of Science and Technology, Wroc{\l}aw, Poland
        \and
             Faculty of Electronics, Wroc{\l}aw University of Science and Technology, Wroc{\l}aw, Poland\\
             }

   \date{Received XXX; accepted XXX}

 
  \abstract
   {Precise continuum normalisation of merged \'{e}chelle spectra is a demanding task necessary for various detailed spectroscopic analyses. Automatic methods have limited effectiveness due to the variety of features present in the spectra of stars. This complexity often leads to the necessity of manual normalisation which is a time demanding task.}
   {The aim of this work is to develop a fully automated normalisation tool that works with order-merged spectra and offers flexible manual fine-tuning, if necessary.}
   {The core of the proposed method uses the novel fully convolutional deep neural network (SUPP Network) that was trained to predict a \textit{pseudo-continuum}. The post-processing step uses smoothing splines that gives access to regressed knots useful for optional manual corrections. The active learning technique was applied to deal with possible biases that may arise from training with synthetic spectra and to extend the applicability of the proposed method to features absent in this kind of spectra.}
   {The developed normalisation method was tested with high-resolution spectra of stars having spectral types from O to G, and gave root mean squared (RMS) error over the set of test stars equal 0.0128 in the spectral range from 3900\,{\AA} to 7000\,{\AA} and 0.0081 in the range from 4200\,{\AA} to 7000\,\AA. Experiments with synthetic spectra give RMS of the order of 0.0050.}
   {The proposed method gives results comparable to careful manual normalisation. Additionally, this approach is general and can be used in other fields of astronomy where background modelling or trend removal is a part of data processing. The algorithm is available online: \protect\url{https://git.io/JqJhf}.}


   \maketitle
%
\section{Introduction}

Electromagnetic spectra of astronomical objects such as stars, galaxies or exoplanets are a very abundant source of information and enable us to study the physics of the objects in detail. Low-resolution spectra are closely related to photometry and can be used to determine basic properties of stars, like effective temperatures, surface gravities or metallicities. High-resolution spectra allow us to investigate profiles of separate absorption and emission lines in detail, and make it possible to measure individual abundances of elements, understand the vertical structure of a stellar atmosphere, and study its velocity fields (e.g. microturbulence, macroturbulence, rotation, stellar oscillations, granulation). 

The continuum-normalisation process is an important part of spectrum preprocessing because usually for the next steps of the analysis the \textit{pseudo-continuum} should be correctly subtracted. The problem of continuum normalisation is non-trivial, as several factors are responsible for the shape of a spectrum, including real stellar continuum shape, the Earth atmosphere, characteristics of optical and electronic components of a spectrograph, presence of cosmic rays, and last but not least, residuals introduced by pipelines used by observatories to calibrate and reduce spectra, due to, for example, imperfect orders merging or blaze function removal. Several existing normalisation methods are worth mentioning, e.g. filtering methods (moving window maximum filtering, asymmetric sigma clipping, etc.) followed by low order polynomial fitting, smoothing in the frequency domain, and methods based on the concept of convex-hull and alpha-shape theories \citep{Xu_2019, Cretignier_2020}. They all try to remove spectral lines from the spectrum and then fit a low-order polynomial or spline function to the remaining \textit{pseudo-continuum}, and they all contain some free parameters that need to be adjusted manually. For a broad overview of methods and problems present in the normalisation of stellar spectra we recommend the article of \citet{Cretignier_2020}.

Normalisation methods based on the polynomial fitting and frequency domain filtering suffer due to the trade-off between the treatment of rapidly changing parts of a continuum (e.g. ripples often present in merged \textit{\'{e}chelle} spectra) and the presence of wide spectral features (e.g. hydrogen lines or molecular bands). These limitations can be partially overcome by introducing some adaptive penalised least-squares terms in the minimisation objective. This approach is extensively used in the field of Raman spectroscopy \citep{Cadusch_2013}. Methods based on the concept of convex hulls also have some limitations. They assume that the local maximum is a good indicator of a stellar continuum. In most cases, this is a reasonable attitude, but breaks for spectra with emission features present and in regions with extensively blended lines, where the continuum is well above the measured spectrum (e.g. in the optical spectra of G or later stars). All these methods share the problem that features located in regions covered with spectral lines are lost. This does not allow for the effective recovery of the \textit{pseudo-continuum} shape in the parts of spectra mentioned above. The continuum in these parts is only smoothed to match the surroundings of the line. Additionally, those methods are particularly sensitive to the level of noise, projected rotational velocity, and resolution of the instrument used to observe the spectra.

These limitations can be overcome when we have a template for a given spectrum. Normalisation can then be reduced to dividing the spectrum by the model and then using the selected trend fitting tool, e.g. adaptive spline functions, to model a \textit{pseudo-continuum}. In this case, the challenge is to successfully find a model for the spectrum, the parameters of which are unknown at the first stage of the analysis. In turn, the proposed approach utilising information from the entire spectrum, also from areas with spectral lines, is resistant to the above-mentioned problems and can be used in a fully automatic manner, offering, at the same time, the flexibility of manual corrections, if necessary.

Nonetheless, an astronomer experienced in spectrum normalisation is able to fit a \textit{pseudo-continuum} taking most of the complexities into account, especially when normalisation is done iteratively during model fitting. One of the tools that make the fully manual continuum fitting and initial stellar parameters' estimation straightforward is the application HANDY\footnote{https://rozanskit.com/HANDY/}. However, manual normalisation has several important drawbacks. It is very time-consuming, prone to human biases, and not reproducible. The manual normalisation cannot be done reliably in regions covered by wide lines and line blends when synthetic spectra are not used as a reference. Manual normalisation strengths suggest that the key points are the understanding of a spectrum and the experience with features that may appear in the \textit{pseudo-continuum}. These findings suggest that machine learning algorithms are promising tools to overcome mentioned limitations.

The use of machine learning algorithms becomes more and more frequent in physical sciences \citep[][]{Carleo_2019}, including astronomy \citep{BALL_2010,baron2019machine}. This progress becomes possible in recent years due to important developments in the algorithms of machine learning, especially in the neural network field, and due to the constant and rapid increase of available computational power. On the other hand, astronomy delivers more numerous and complex astrometric \citep[e.g. Gaia mission][]{2016A&A...595A...1G}, photometric \citep[e.g. Zwicky Transient Facility and future Vera Rubin Observatory with Legacy Survey of Space and Time][]{2019PASP..131c8002M, Ivezi_2019} and spectroscopic (e.g. ESO database\footnote{https://archive.eso.org/cms/data-portal.html}; APOGEE, \citet{2017AJ....154...94M}; LAMOST, \citet{zhao2012lamost}) databases, which grow in size rapidly. Machine learning was applied for various tasks in the field of astronomy, e.g. real-time detection of gravitational waves and parameter estimation \citep{George_2018}, estimation of effective temperatures and metallicities of M-type stars \citep{antoniadis2020}, estimation of initial parameters for asteroseismic modelling \citep{2019PASP..131j8001H}, classification of diffuse interstellar bands \citep{2019PASP..131j8001H} and morphological segmentation of galaxies \citep{FARIAS2020100420}.

We explore deep artificial neural networks (DNN) in search of architectures that can deal with a \textit{pseudo-continuum} prediction task. We propose a new algorithm, based on the neural network SUPPNet, that achieves results comparable to human professionals.


\section{Machine Learning and spectrum normalisation}

Up to our knowledge, this is the first approach towards automated stellar spectrum normalisation using tools rooted in the deep learning field. From the above examples, it is clear that fundamental limitations of the previous methods are: an implicit assumption that a local maximum in a spectrum is a good indicator of a \textit{pseudo-continuum}; the necessity of manual adjustments of several parameters; often poor results when the \textit{pseudo-continuum} passes trough emission features; ripples that arise from orders merging (\'{e}chelle spectra). Deep neural networks are natural candidates to overcome these limitations, as in principle they are capable to grasp complex priors and approximate any function. It means that a suitable neural network model after training on a representative dataset should learn to recognise a type of spectrum and a \textit{pseudo-continuum} and be able to fit the \textit{pseudo-continuum} correctly in most cases. In this approach, the quality of a spectrum normalisation algorithm is restricted by the generality and quality of an available training set.

To fit the normalisation problem in the deep learning framework we consider spectrum normalisation as filtering from the domain of spectrum measurement to the domain of possible \textit{pseudo-continua}. As a filter, that is flexible enough to implement such a mapping, we used a fully convolutional neural network of novel architecture, which is mainly based on the work done in the field of computer vision, in particular in the semantic segmentation problem.


\subsection{Training data description}
A machine learning model can be as good as data used for its training but not any better. Because of that, much attention was paid to preparing the training data properly. We applied the active learning technique so we prepared two datasets: the first, composed of synthetic spectra only, and the second based on automatically normalised and manually corrected observational spectra. In this work, we denote supplementation of the training process with spectra and \textit{pseudo-continua} from manual normalisation as active machine learning. 

We started by preparing an extensive set of synthetic continuum-normalised spectra. We used SYNTHE/ATLAS \citep{kurucz1970atlas} codes to compute 10000 spectra with randomly selected atmospheric parameters (effective temperature, $T_\textrm{eff}$, ranging from 3000\,K to 30000\,K and logarithm of surface gravity, $\log g$, from 1.0 to 5.5). We also used BSTAR and OSTAR grids \citep{2003ApJS..146..417L,2007ApJS..169...83L}, which together span effective temperature range from 15000\,K to 55000\,K, and logarithm of surface gravity from 1.75 to 4.75 (not in the whole range of temperatures). The main difference between these two sources of spectra is a different treatment of atomic levels populations. The ATLAS code solves the classical stellar atmosphere problem, and SYNTHE computes the synthetic spectrum assuming local thermodynamic equilibrium (LTE), which means that level populations are calculated using the Boltzmann distribution and the Saha ionisation equation. In the case of BSTAR and OSTAR grids, they were calculated using SYNSPEC/TLUSTY codes that explicitly solve rate equations for a chosen set of levels. This is especially important for hotter stars, where the radiative processes dominate the collisional transitions and non-LTE effects become prominent. Including non-LTE effects leads to changes of line depths and, what is more important, frequently results in the appearance of emission features.

Some families of analytical functions (sinusoidal, smoothed sawtooth, Akima spline with 5 knots) and also some continua coming from manual normalisation were used as artificial \textit{pseudo-continua} shapes. To build the first training dataset the spectrum from the set of prepared synthetic spectra was sampled, part of it was randomly chosen. Next, it was convolved with a random broadening kernel and finally multiplied by an artificial continuum. This procedure repeated about 150000 times gave a large set composed of diverse training samples. The broadening kernel included rotation, macroturbulence and instrumental broadening. Rotation and macroturbulence velocities were chosen randomly in physically reasonable ranges. A projected rotation velocity ($v\sin i$) range depends on effective temperature and is drawn from a uniform distribution with the lower boundary equal to zero, and the upper one successively equal to 50, 100, 200, 300, 400\,km\,s$^{-1}$ in the ranges (< 5000), (5000, 6000), (6000, 7500), (7500, 10000) and (>10000)\,K, respectively. This distribution is based on the discussion presented by \citet{2009LNP...765..207R}. The macroturbulence velocity, $\zeta_v$, was drawn from the same uniform distribution in the range from 0 to 30\,km\,s$^{-1}$, regardless of effective temperature. The chosen instrumental resolution (from 40000 to 120000) covers a range of the most available high-resolution spectrographs.

The second stage of learning used the training set extended by the active dataset, which is composed of automatically normalised and manually corrected observational spectra \citep[application and description of active learning in astronomical context can be found in the work by][]{2020A&A...643A.122S}. First, the model trained on synthetic data was used to normalise spectra from UVES Paranal Observatory Project \citep[UVES POP\footnote{https://www.eso.org/sci/observing/tools/uvespop.html};][]{bagnulo2003uves} (IC\,2391, NGC\,6475, and the brightest stars of southern sky, resolution equal 80000), and the set of FEROS\footnote{https://www.eso.org/sci/facilities/lasilla/instruments/feros.html} spectra (153 spectra with $\textrm{SNR}>500$ without object duplicates, resolution equal 48000). Then normalisation was manually checked and carefully corrected for each automatically processed spectrum. That resulted in a set of around 250 normalised spectra and \textit{pseudo-continua} fits, making up the active dataset.


\subsection{Tasks description}

\begin{figure}
\centering
\includegraphics[width=\hsize]{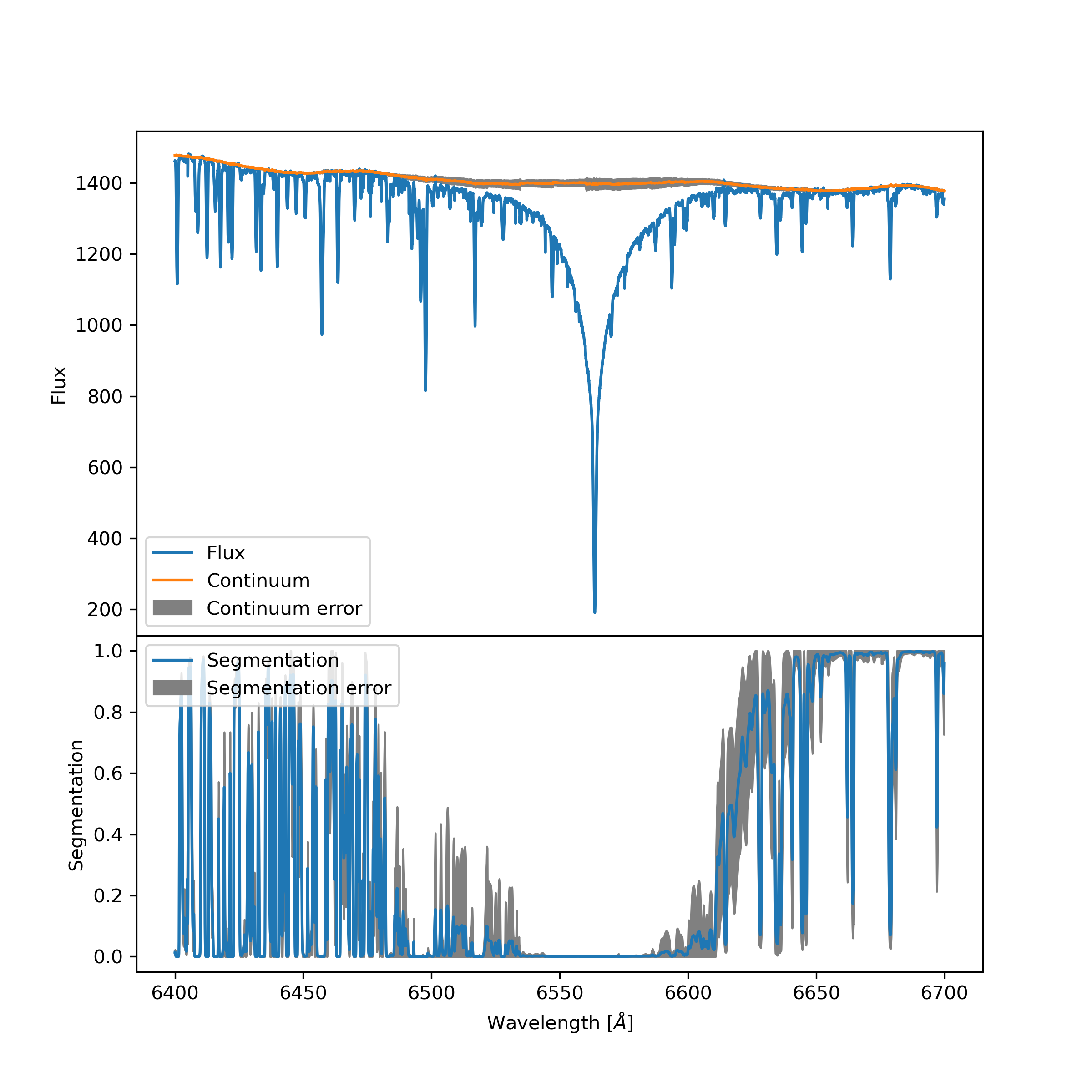}
  \caption{Result of HD\,27411 (A3m) spectrum processing. The left upper panel shows the spectrum and the predicted continuum near H$\alpha$ line. The left lower panel shows the corresponding segmentation mask. The shaded area denotes the estimated uncertainty.
          }
     \label{fig:example_result}
\end{figure}

Initial tests of deep learning methods in stellar spectrum processing focused on two distinct tasks: segmentation of spectrum into \textit{pseudo-continuum} and \textit{non-pseudo-continuum} parts, and \textit{pseudo-continuum} prediction. The segmentation can be considered as a subclass of a classification problem. It aims to predict the class that a segment of input data belongs to. The segment can be as small as one pixel in the case of image segmentation or flux measurement at a given wavelength (sample) in the case of one-dimensional spectral data segmentation. In our case we classify spectrum sample-wise into two classes: "continuum", and "non-continuum" (see the bottom panel of Fig. \ref{fig:example_result} for an example). 

It can be given by a function $f:~\mathcal{R}^{n} \rightarrow \{0,1\}^n$, where $n \in \mathcal{N}$, is the number of samples in the input and the output, $1$ corresponds to the \textit{pseudo-continuum} class, and $0$ to the \textit{non-pseudo-continuum} class. It is important to note that the result returned by the last layer of a neural network is composed of real numbers in the range from 0 to 1. Thresholding must be applied to obtain discrete classes, . The common choice used also in this work is $0.5$. Spectrum segmentation was considered here as an auxiliary task and as a potential regularisation technique \citep{2017arXiv171010686K}.

A functional form of continuum prediction, which is a multidimensional regression, differs slightly from the segmentation and is given by $ f:~\mathcal{R}^{n} \rightarrow \mathcal{R}^n $, with the same notation as above. The co-domain is here over the real numbers, instead of a discrete set of values that represent different classes. An exemplary result of such a prediction for both, segmentation and a \textit{pseudo-continuum} prediction task is given in Fig. \ref{fig:example_result}.

\subsection{Loss functions, metrics and optimisers}

A neural network learning needs an optimisation algorithm that adjusts well free parameters of the model to the approximate trained function. The quality of the obtained mapping can be evaluated using various loss functions and metrics. A loss function denotes a differentiable function whose value is optimised during the training process by the algorithm that iteratively adjusts free parameters of a neural network under the training. A metric is a function that is used to evaluate the performance of the model but is not directly used for optimisation during the training. Metric functions can be non-differentiable. The adopted naming convention comes from the TensorFlow machine learning library \citep{tensorflow2015-whitepaper}.

The \textit{cross-entropy} loss function is often used for segmentation:

$$
J(y,\hat{y})= \sum_{i=0}^N H(y_i,\hat{y}_i) = \sum_{i=0}^N \sum_j y_{i,j}\log \hat{y}_{i,j},
$$ 
where $N$ is a number of samples, $y_i$ is a vector of target classes for i-th sample, and $\hat{y}_i$ is a vector of the model's output for the same sample. $y_{i,j}$ equals $1$ only if i-th sample belongs to j-th class, so $y_i \in \{0,1\}^M$, where M is a number of classes. The cross-entropy equals zero if a model perfectly assigns classes to samples. In this work, the \textit{binary-cross-entropy} loss was applied as there are only two distinct classes. The \textit{accuracy}, which is given by the number of correct predictions divided by the number of all predictions was used as a metric. Here, in the case of the binary segmentation, the $accuracy = (\textrm{TP} + \textrm{TN})/N$, where TP denotes the number of true positive predictions, TN corresponds to true negative, and $N$ gives the number of samples.

The \textit{mean squared error} (MSE) was used as a loss function for the \textit{pseudo-continuum} prediction. It is given by:

$$
\textrm{MSE}(y,\hat{y})= \frac{1}{N}\sum_{i=0}^N (y_i-\hat{y}_i)^2,
$$
where $N$ denotes the number of samples, $y$ is a vector of target values, and $\hat{y}$ is an estimated vector. MSE is a standard loss function for a regression problem. The \textit{mean absolute error} (MAE) was used as a metric for this task. The functional form of MAE is given by: 
$$
\textrm{MAE}(y,\hat{y})= \frac{1}{N}\sum_{i=0}^N |y_i-\hat{y}_i|.
$$

All performed experiments use the Nesterov-accelerated Adaptive Moment Estimation optimiser \citep[Nadam;][]{dozat2016incorporating}, which iteratively adjusts free parameters of a neural network during a training process. It is based on two main concepts. The first is \textit{Nesterov's momentum} which is beneficial in dimensions of an objective function with small curvature that consistently points one direction \citep{nesterov1983method}. The second concept is the adaptive learning rate that is able to compute individual learning rates based on the first and the second moments of gradients \citep[Adam;][]{kingma2017adam}, and works well with sparse gradients and non-stationary objectives.

\subsection{Exploratory neural network tests}
\label{sec:exploratory_tests}
Initial exploration of neural network architectures was focused on tasks of \textit{pseudo-continuum} prediction and spectrum segmentation. As these problems are inherently one-dimensional (inputs and targets are sequential data), we have implemented and tested one-dimensional versions of several neural network architectures that were successfully used in the field of image segmentation. We expected that advances made in the two-dimensional domain, with necessary adaptations and some minor changes, could be adopted for the processing of one-dimensional signals that stellar spectra are. The purpose of these tests was to check this hypothesis. To some extent, the method to find the best neural network follows the work of \citet{2020arXiv200313678R}.

The selection of potential architectures was based in particular on the paper of \citet{Hoeser_2020}, which gives a detailed overview of both, the historical development and current state-of-the-art solutions. The architectures selected for the experiments were: Fully Convolutional Network \citep[FCN,][]{long2015fully}, Deconvolution Network \citep[DeconvNet,][]{noh2015learning}, U-Net \citep{ronneberger2015unet}, UNet++, \citep{zhou2018unet}, Feature Pyramid Network \citep[FPN,][]{lin2017feature, kirillov2019panoptic}, and Pyramid Scene Parsing Network \citep[PSPNet,][]{zhao2017pyramid}. Additionally, a new architecture -- UPPNet, which is an extension of the U-Net architecture was proposed.

The same training and evaluation scheme was used for all tests to make the comparison informative. Each tested neural network is composed of a body and a prediction head. The body is the part sampled from the adopted parameters space, considering the type of architecture and its free hyper-parameters (e.g. a number of layers, etc.). As an input, it takes a one-dimensional spectrum (vector of the length equals 8192) and outputs several feature maps of the same length as the input spectrum (e.g. in the case of a body that return 8 feature maps, a matrix of size (8192,8) is an output). In the context of a one-dimensional spectrum, feature maps can represent different shapes of spectral lines. Schematic diagrams of all tested bodies can be found in Figs.\,\ref{fig:all_nets-FCNet}-\ref{fig:all_nets-UPPNet} in Appendix \ref{appendix:diagrams_of_nn}. As the main building block, all networks use residual bottleneck blocks with group convolution \citep{2016arXiv161105431X}, referred here as residual blocks (RB). Hyper-parameters of RBs are: a number of feature maps, a group width and a bottleneck ratio (in all experiments the bottleneck ratio was fixed to one, see \citet{2020arXiv200313678R}). The head takes feature maps from the body part as an input and returns a final prediction. All heads share the same architecture, i.e. have three point-wise convolutional layers \citep{lecun_1989} with the number of features equal 64, 32 followed by 1. In the first two layers the ReLU ($\textrm{ReLU}(x)=\max(x,0)$) activation function was used, while in the third one, it was the softmax function for segmentation and ReLU for \textit{pseudo-continuum} prediction. The exception is the Fully Convolutional Network, whose architecture does not allow for this consistent approach. In this case, we skipped some of the tests (details are given in the architecture description below).

\begin{figure}
\centering
\includegraphics[width=\hsize]{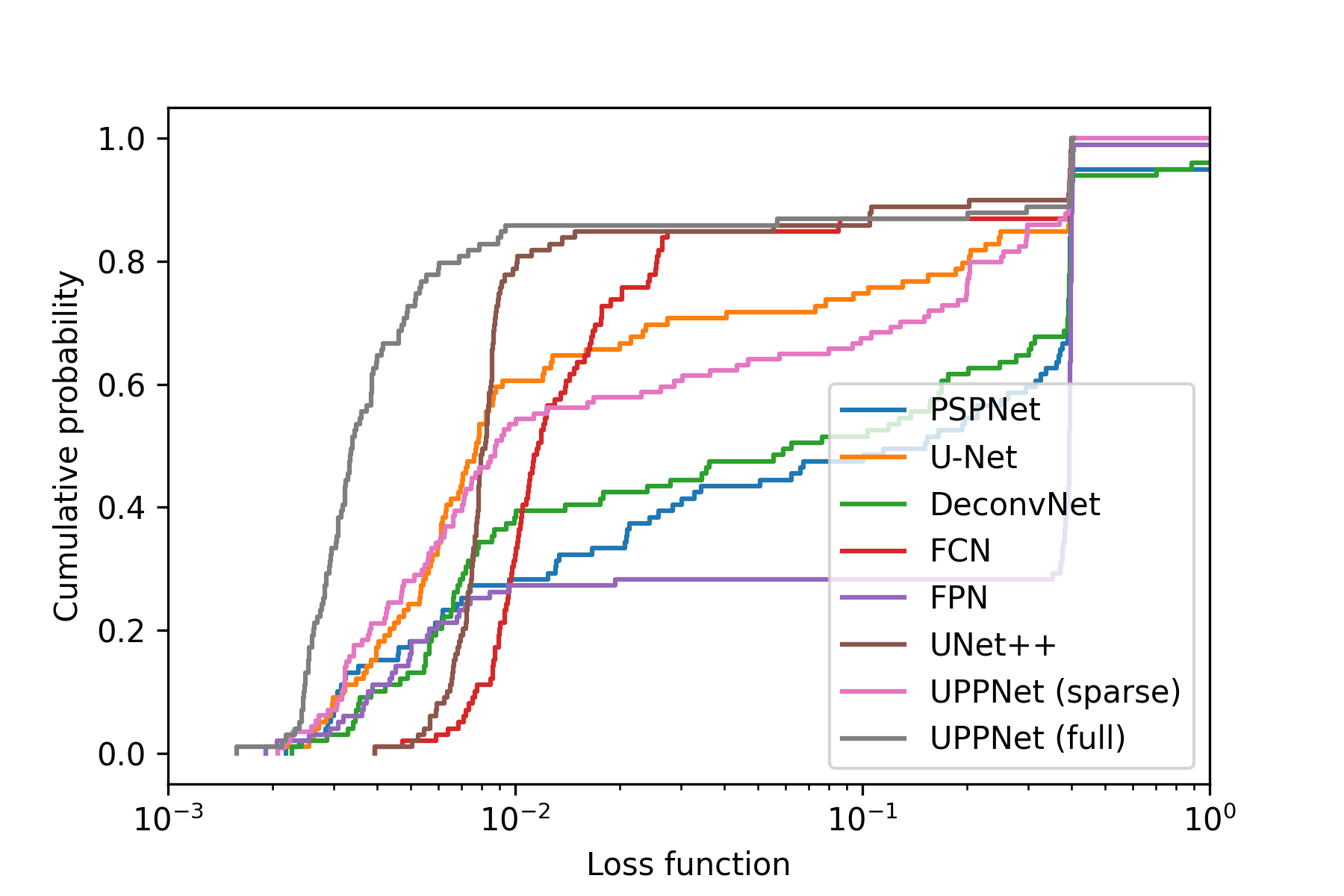}
  \caption{Distribution of loss function values of neural networks randomly sampled from tested architectures trained in the task of \textit{pseudo-continuum} prediction. One hundred random neural networks, with a number of trainable parameters ranging from 200000 to 300000, were drawn for each architecture. The training was held in a low-data regime, for only 30 epochs.}
     \label{fig:models_error_distribution}
\end{figure}

One hundred neural architecture realisations with a number of free parameters ranging from 200000 to 300000 were sampled from each architecture and trained in a low-data regime ($\sim$5000 training samples from the synthetic training dataset, 30 epochs training, where epoch means single complete pass through the training data) in a \textit{pseudo-continuum} prediction task. Networks' validation was performed on the validation dataset separated at the beginning from the training set. The obtained distribution of models that can be found in Fig.\,\ref{fig:models_error_distribution}, which shows the probability that the loss value on a validation set of a neural network sampled from a given architecture will be lower than a corresponding abscissa value. For example, approximately 60 percent of neural networks of the UPPNet (full) architecture have a loss function value lower than $4\times10^{-3}$. This plot shows that neural networks of different architectures differ in the concentration of high-quality models in their parameters space and in the mean squared error of the best neural network. The UPPNet (full) can be thus considered the most promising architecture as it gives many relatively good models and also has the best model among all the trained networks.

In the second step of exploration, the best neural network from each architecture was picked and trained in segmentation/\textit{pseudo-continuum} prediction, and also in both tasks simultaneously. In the case of the simultaneous training neural networks were equipped with two independent prediction heads. This training covered the entire synthetic dataset and lasted 150 epochs, where epoch means a single complete pass through the training data. The metrics were calculated on the same validation dataset as in the first step of exploration. The first 100 epochs used a learning rate equal $10^{-4}$ and a learning rate equal $10^{-5}$ was used for the remaining steps. 

Brief summary of the results can be found in Table\,\ref{tab:architectures_experiments}. Mean absolute error or accuracy is shown for each architecture and task. The best value in each column is given in bold. UPPNet (full), trained on both tasks simultaneously is leading in \textit{pseudo-continuum} prediction with MAE equal 0.0110, while in segmentation the best is UPPNet (sparse) trained only in the segmentation task, with accuracy equal 0.9166. The best, when trained in \textit{pseudo-continuum} prediction only, is the U-Net network, with MAE equal to 0.0112. Training in both tasks simultaneously results in a systematic decrease in the segmentation quality but has little effect on the \textit{pseudo-continuum} estimation. An overview of original models used in the image semantic segmentation, with a short note about their one-dimensional versions with the results and the conclusions is provided below.

\begin{table}
\caption{Results of experiments with various neural networks architectures on the validation dataset. S stands for segmentation, C for \textit{pseudo-continuum} prediction. We report accuracy metrics for the former and mean absolute error for the latter. The last two columns contain the results achieved by models trained in both tasks simultaneously (C\&S). The best results are in bold. The results in the last row anticipate the metrics of the final model. These models are described in detail in Section \ref{sec:SUPP_Network}.}
\label{tab:architectures_experiments}
\centering
\begin{tabular}{lllll}
\hline\hline
\multicolumn{1}{c}{Architecture} & \multicolumn{1}{c}{C} & \multicolumn{1}{c}{S} & \multicolumn{2}{c}{C\&S}                            \\ 
\multicolumn{1}{c}{}                            & \multicolumn{1}{c}{}             & \multicolumn{1}{c}{}                 & \multicolumn{1}{c}{C} & \multicolumn{1}{c}{S} \\ 
\hline
FCN                                             & 0.0580                              & 0.8877                                  & --                              & --              \\
DeconvNet                                       & 0.0124                              & 0.8991                                  & 0.0128                              & 0.8763       \\
U-Net                                           & \textbf{0.0112}                     & 0.9105                                  & 0.0119                              & 0.8866        \\
UNet++                                          & 0.0146                              & 0.8945                                  & 0.0133                              & 0.8679          \\
FPN                                             & 0.0125                              & 0.9063                                  & 0.0119                              & 0.8887           \\
PSPNet                                          & 0.0115                              & 0.9154                                  & 0.0119                              & 0.8914  \\
UPPNet (sparse)                                   & 0.0122                              & \textbf{0.9166}                         & 0.0126                              & 0.8857           \\
UPPNet (full)                                     & 0.0116                              & 0.9065                                  & \textbf{0.0110}            & \textbf{0.9033}          \\
\hline
SUPPNet (synth)                                    & -                              & -                                  & \textbf{0.0092}            & \textbf{0.9132}          \\
\hline
\end{tabular}
\end{table}

\subsubsection{Fully Convolutional Network}
This type of convolutional neural network (CNN) was proposed by \citet[][]{long2015fully} and influenced the whole field of the image semantic segmentation -- most of the following architectures used in the semantic segmentation are also fully convolutional neural networks. This end-to-end trainable model uses a VGG-16 network \citep{simonyan2015deep} as a backbone. This is a simple network that contains 16 trainable layers. The segmentation prediction is a function of feature maps previously upsampled by trained deconvolution layers from several intermediate representations, and merged by summation. This design gives spatial resolution and access to high level semantic classes. In the context of one-dimensional spectrum processing, high level semantic information can for example correspond to the presence of emission or wide absorption features in a spectrum.  In all the following networks it is always the main point: to transfer somehow to the final segmentation result the information about precise localisation that the input image contains, but also to discover the class, that complex objects present in the input belong to. The schematic diagram of this architecture can be found in Appendix \ref{appendix:diagrams_of_nn} in Fig.\,\ref{fig:all_nets-FCNet}.

FCN architecture does not fit the comparison strategy used in our work, but we include it here as a prototype for all fully convolutional neural networks. It forms the low-resolution prediction, gradually refines it, and increases its resolution by upsampling and addition. This is contrary to other network architectures, that form high resolution features used by a neural network head for the final prediction. Due to these characteristics, the FCN network that predicts \textit{pseudo-continuum} and segmentation simultaneously was not implemented.

In the context of stellar spectra processing, this network has the worst results among all tested architectures. MAE of the predicted \textit{pseudo-continuum} equals 0.0580 and the accuracy of segmentation is 0.8877. It shows that strategy of upsampling low-resolution prediction is not the efficient approach to the investigated problems.

\subsubsection{Deconvolution Network}
This network \citep{noh2015learning} is an example of a so called encoder-decoder architecture. It is composed of two parts, encoder -- narrowing part, which is the VGG-16 network, and decoder -- widening part, the mirrored VGG-16 where pooling is replaced with unpooling layers that use spatial positions of elements pooled in the decoder. This means the localisation information is forwarded to the final segmentation output. The schematic diagram of this architecture can be found in Appendix\,\ref{appendix:diagrams_of_nn} in Fig.\,\ref{fig:all_nets-DeconvNet}.

One-dimensional version of this architecture gives moderate results (separate training: $\textrm{MAE} = 0.0124$,  $\textrm{accuracy} = 0.8991$). Training in both tasks decreases prediction quality (C\&S training: $\textrm{MAE} = 0.0128$,  $\textrm{accuracy} = 0.8763$). To accurately preserve spatial location information, which is a key to accurately predict \textit{pseudo-continuum} and segmentation, is the challenge with this architecture. We believe that this is the reason for the moderate performance of this network.

\subsubsection{U-Net}
The U-Net architecture proposed by \citet{ronneberger2015unet} is a very successful and broadly exploited model. It is another encoder-decoder network. Its encoder is composed of five convolutional blocks that double the number of feature maps at each stage, and that are separated by maximum pooling layer \citep[comparison of pooling methods can be found in][]{scherer2010evaluation} with $2\times2$ size and stride. The decoder is also composed of five blocks and upsamples the low resolution -- high semantic level feature maps, but in contrary to Deconvolution Network, blocks do not use only pooling indices but also concatenate the encoder feature map of the corresponding resolution. This is an example of a widely used concept of skipped connections, that helps to propagate the gradient in the training process, and helps the network to recover spatial localisation of objects in the input image. The blocks use $3\times3$ convolution kernels, except the final layer which uses $1\times1$ convolution for the final prediction. The schematic diagram of this architecture can be found in Appendix \ref{appendix:diagrams_of_nn} in Fig.\,\ref{fig:all_nets-U-Net}.

Tested implementation of one-dimensional U-Net architecture gives very good results in both segmentation and \textit{pseudo-continuum} prediction (separate training: $\textrm{MAE} = 0.0112$,  $\textrm{accuracy} = 0.9105$). Its \textit{pseudo-continuum} prediction MAE is the lowest among networks with one output. Simultaneous training to predict both targets does not lead to any improvements (C\&S training: $\textrm{MAE} = 0.0119$,  $\textrm{accuracy} = 0.8866$). This simple architecture is a very strong baseline for any further experiments.

\subsubsection{UNet++}
UNet++ \citep{zhou2018unet}, is an extension of U-Net architecture. It replaces the skipped connections with densely connected blocks and uses deep supervision for regularisation. The authors of the original article argue that the proposed connection scheme bridges a semantic gap between feature maps obtained in encoder and decoder part. The schematic diagram of this architecture can be found in Appendix \ref{appendix:diagrams_of_nn} in Fig.\,\ref{fig:all_nets-UNetpp}.

In the original work, the authors used intermediate supervision, but we do not use it for the sake of fair comparison to other architectures. Although in theory UNet++ is superior to simpler U-Net architecture, tested one-dimensional UNet++ network shows poor results (separate training: $\textrm{MAE} = 0.0146$,  $\textrm{accuracy} = 0.8945$). Training in both tasks simultaneously slightly improves \textit{pseudo-continuum} quality ($\textrm{MAE} = 0.0133$) but degrades segmentation substantially ($\textrm{accuracy} = 0.8679$). Nonetheless, we suspect that this architecture could outperform U-Net in the regime of bigger networks, where the U-Net performance could potentially saturate.

\subsubsection{Feature Pyramid Network}
Feature Pyramid Network (FPN) was originally developed as a backbone for a two-stage object detection model and later was used in a panoptic segmentation task \citep[the task that unifies image semantic and instance segmentation.,][]{kirillov2019panoptic}. The basic idea is to incorporate the construction of a feature pyramid as a part of neural network architecture. The feature pyramid is implemented as a path that upsamples the features using the nearest neighbour interpolation and uses lateral connections for better spatial localisation of high-resolution feature maps. It is visually similar to the U-Net architecture but conceptually implements a different idea. FPN propagates the same feature maps across different resolutions, while U-Net in principle may alter the feature maps towards the network output.  The schematic diagram of this architecture can be found in Appendix \ref{appendix:diagrams_of_nn} in Fig.\,\ref{fig:all_nets-FPNet}.

FPN gives promising results in spectra processing (separate training: $\textrm{MAE} = 0.0125$,  $\textrm{accuracy} = 0.9063$). Both-task training slightly improves the metrics of \textit{pseudo-continuum} prediction ($\textrm{MAE} = 0.0119$) but worsens segmentation quality ($\textrm{accuracy} = 0.8887$).

\subsubsection{Pyramid Scene Parsing Network}
Pyramid Scene Parsing Network (PSPNet) \citep{zhao2017pyramid} is composed of three parts: a backbone network that delivers a feature map, Pyramid Pooling Module (PPM) that helps to introduce contextual information from different parts of an image to the final prediction, and an output convolutional network that is responsible for the final prediction. Usage of the PPM that pools the features on different scales, applies $1\times1$ convolution, bilinear upsampling, and concatenate produced features to input the feature map of the module is the novelty. The authors experimentally show that such a lightweight module is able to introduce contextual information in the final prediction. The schematic diagram of this architecture can be found in Appendix\,\ref{appendix:diagrams_of_nn} in Fig.\,\ref{fig:all_nets-PSPNet}.

\begin{figure}
\resizebox{\hsize}{!}
{\includegraphics[clip]{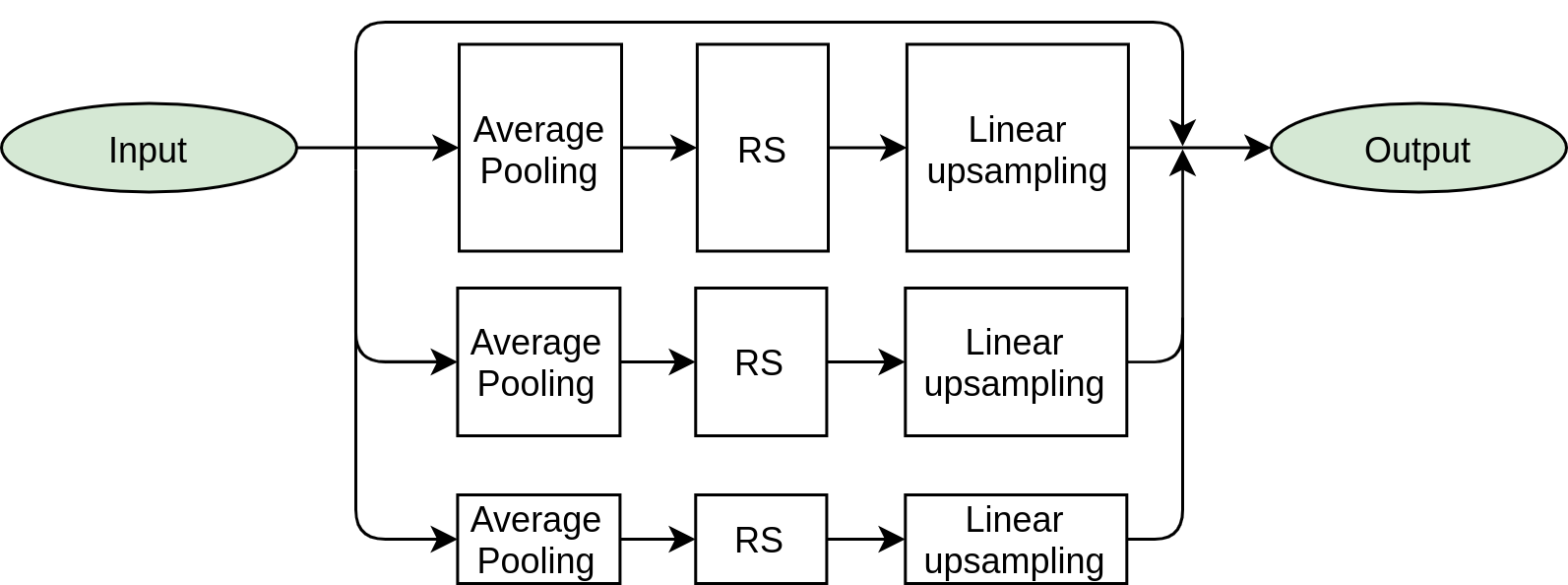}}
\caption{Pyramid Pooling Module (PPM) used in PSPNet and UPPNet networks. The PPM pools input feature maps at different scales, processes them using residual stages, linearly upsamples features to input resolution and finally concatenates them with input features. In this work, the PPM pools the input features to all resolutions that strictly divide the input resolution, e.g. for features' resolution equal 32 it pools at scales: 2, 4, 8, and 16. The number of residual blocks in each RS and the number of features in each residual block were the same for all PPMs used in exploratory tests and equal respectively 4 and 8.}
\label{fig:all_nets-PPM}
\end{figure}

One dimensional version of PSPNet uses PPM depicted schematically in Fig.\,\ref{fig:all_nets-PPM}. Although PSPNet does not use skipped connections in between its encoder and decoder, it gives great results (separate training: $\textrm{MAE} = 0.0115$,  $\textrm{accuracy} = 0.9154$). Its quality slightly degrades when trained on both tasks (C\&S training: $\textrm{MAE} = 0.0119$,  $\textrm{accuracy} = 0.8914$). Its results in \textit{pseudo-continuum} prediction are very close to results of the U-Net, while in segmentation it achieves results about 0.5\% better in accuracy. 

\subsection{UPPNet}
\label{sec:UPPNet}
\begin{figure}
\resizebox{\hsize}{!}
{\includegraphics[clip]{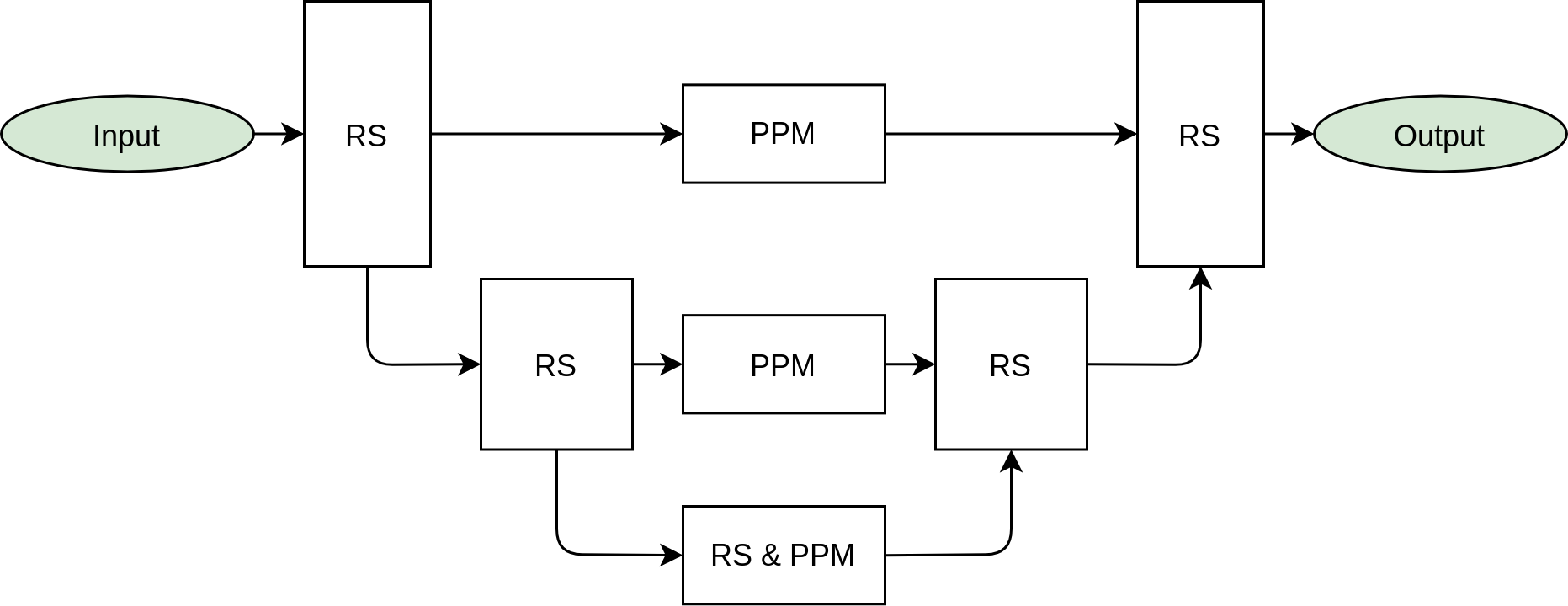}}
\caption{Diagram of U-Net with Pyramid Pooling Modules -- UPPNet. The two residual stages (RS) on the left create the narrowing path. The downward arrows represent strided residual blocks that decrease the sequence length by the factor of two. The central part has three PPM modules, the bottom one is preceded by RS. The widening path, on the right, is a reflection of the narrowing path. Upward arrows represent upsampling by the factor of two. The upsampled features are concatenated with the result from the PPM blocks before being fed into the RS blocks. The depth of this UPPNet is defined as two.}
\label{fig:UPPNet_text}
\end{figure}

Insights from the previously mentioned architectures have encouraged us to experiment with a different placement of skipped connections together with the use of Pyramid Pooling Modules. This has led to U-Net with Pyramid Pooling Modules (UPPNet). First, we hypothesised that not all skipped connections are needed, but at the same time we suspected that PPMs included at some depths may result in better predictions. The proposed architecture is derived from U-Net by randomly dropping skipped connections or by replacing them with PPM. This version of UPPNet is denoted as \textit{sparse}. U-Net, DeconvNet, and PSPNet are special cases of this architecture. Next, we decided to experiment with a version of UPPNet which is derived from the U-Net by replacing all skipped connections with PPMs, with an additional PPM module at the bottom of the network. This version of UPPNet is denoted as \textit{full}. The schematic diagram of UPPNet (full) can be found in Fig.\,\ref{fig:UPPNet_text}. 

Figure\,\ref{fig:models_error_distribution} partially justifies the idea behind sparse UPPNet as there are relatively many networks giving loss values lower that $4\times10^{-3}$, but Table \ref{tab:architectures_experiments} shows that in full training regime this additional degree of freedom in skipped connections arrangement not necessarily leads to results better than obtained with U-Net and PSPNet (separate training: $\textrm{MAE} = 0.0122$,  $\textrm{accuracy} = 0.9166$, C\&S training: $\textrm{MAE} = 0.0126$,  $\textrm{accuracy} = 0.8857$).

The regular UPPNet (full) architecture generally gives better results. Figure\,\ref{fig:models_error_distribution} shows that there are relatively many models with low loss value and that the best single network is of this kind. In the training on the full synthetic dataset (see Table \ref{tab:architectures_experiments}) the advantage of this network decreases and the quality of \textit{pseudo-continuum} prediction is comparable to those of U-Net and PSPNet models (separate training: $\textrm{MAE} = 0.0116$, C\&S training: $\textrm{MAE} = 0.0110$). Nonetheless, the best result in \textit{pseudo-continuum} normalisation belongs to UPPNet (full) architecture when trained in both tasks. Additionally, visual inspection of the results of U-Net and UPPNet showed that the latter gives slightly smaller residuals on hydrogen H$\alpha$ and H$\beta$ spectral lines which are important for atmospheric parameters' estimation. The quality of segmentation is moderate in the case of training only in this task ($\textrm{accuracy}=0.9065$), but is the best among models trained in both tasks simultaneously ($\textrm{accuracy}=0.9033$). Because of these findings, the final model for \textit{pseudo-continuum} prediction uses UPPNet (full) architecture as its main building block and is trained in both tasks simultaneously.

\section{SUPP Network}
\label{sec:SUPP_Network}
Stacked U-Net with Pyramid Pooling Modules (SUPPNet) is a proposed neural network that gives the best results in the \textit{pseudo-continuum} prediction task (see Table\,\ref{tab:architectures_experiments}). This neural network was inspired by the three following well-known solutions present in machine learning literature: the U-Net architecture \citep{ronneberger2015unet}, which is a basic block effective in tasks that combines precise localisation of features with complex semantic concepts, the Pyramid Scene Parsing Module \citep{zhao2017pyramid}, which enables more effective share of contextual information across the whole receptive field, and the Stacked Hourglass Network \citep{newell2016stacked}, for which repetitive bottom-up processing, in conjunction with deep supervision, allows the network to learn fine-grained predictions. 

\begin{figure}
\resizebox{\hsize}{!}
{\includegraphics[clip]{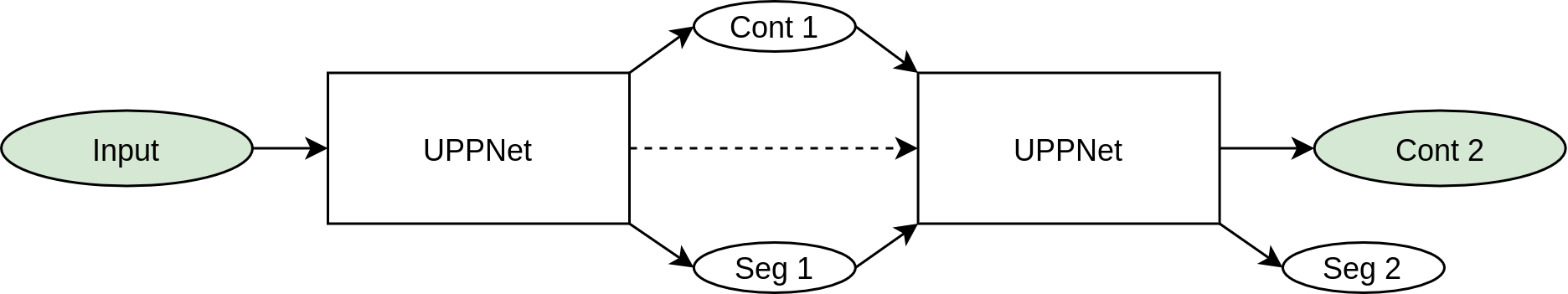}}
\caption{Block diagram of the proposed SUPP Network. The network is composed of two UPPNet blocks and four prediction heads. The first UPPNet block forms coarse predictions, and high resolution features maps that are forwarded to the second block (dashed arrow). The coarse predictions in intermediate outputs \textit{Cont 1} and \textit{Seg 1} (first \textit{pseudo-continuum} and segmentation outputs, respectively) are forwarded to the second block. The second block forms the final predictions at \textit{Cont 2} and \textit{Seg 2} outputs. }
\label{fig:SUPP_Net}
\end{figure}

\subsection{Architecture details}

SUPPNet is composed of two UPPNet (full) blocks and four prediction heads (Seg\,1, Seg\,2, Cont\,1 and Cont\,2; see Fig.\,\ref{fig:SUPP_Net}). The UPPNet module was chosen as the main building block because of its high quality reviewed in Section \ref{sec:UPPNet}. The heads compute final predictions from the high-resolution feature maps created by UPPNet blocks and have simple architecture, described in Section \ref{sec:exploratory_tests}. The first block is responsible for the coarse prediction of a continuum/non-continuum segmentation mask (Seg\,1) and \textit{pseudo-continuum} (Cont\,1). The mentioned outputs are used only while training, for deep supervision (intermediate supervision) \citep{newell2016stacked, zhou2018unet}, which is beneficial in two ways. First, it helps to deal with the vanishing gradient problem. Second, the deep supervision regularises the neural network model. As an input, the second block takes spectrum, intermediate predictions (Seg\,1 and Cont\,1) and high-resolution feature maps from the first UPPNet block concatenated together. It outputs improved features that are used by the Seg\,2 and Cont\,2 heads for final predictions.

Both UPPNet blocks shared hyper-parameters (e.g. a depth, a number of residual blocks in each residual stage, etc.) but not their weights. The choice of hyper-parameters was based on a sampling of 400 UPPNet models that were trained in a low-data regime in a \textit{pseudo-continuum} prediction task, following the procedure from Section \ref{sec:exploratory_tests}. Subsequently, the best UPPNet was selected. The depth of this model equals 8 (its narrowing path has 8 pooling layers in total), with the number of residual blocks in residual stages successively equal 1, 1, 1, 2, 2, 5, 6, 7, 10, 7, 6, 5, 2, 2, 1, 1, 1 with the number of filters at each stage equal respectively 12, 16, 16, 20, 24, 32, 44, 44, 44, 44, 44, 32, 24, 20, 16, 16, 12. The narrowing and widening parts are symmetrical with respect to the bottom residual stage. PPM modules in SUPPNet have four filters and use residual stages that are composed of a single residual block. Their group width equals 64 so residual stages effectively use non-grouped convolutions. The total number of free parameters of the proposed SUPPNet approximately equals one million.

\subsection{Training details}
First, the SUPPNet model was trained on a synthetic dataset only. The training data was fed into the model using a data generator that augmented training examples by randomly cropping the spectra with ratio ranging from 0.7 to 2.5, reflecting the spectrum over the y-axis with 0.5 probability, and adding Gaussian noise to obtain a signal-to-noise ratio from 30 to 500. The ratio equals one means that an input sampling of 0.03\,\AA~was kept. The ratio 0.7 corresponds to a sampling 0.021\,{\AA} and a spectral range $170$\,{\AA} ($170\approx 0.021\times8192$), and 2.5 to a sampling of 0.075\,\AA~ that corresponds to an $610$\,\AA input window width. In this setting, the average sampling of the training data was equal $0.05$\,\AA and this is the value used later for prediction. 

The Nadam optimiser was used for training as in all other experiments described in this work. A learning rate (\textit{lr}) equals $10^{-4}$ for the first 100 epochs and was decreased to $10^{-5}$ for the remaining 50 epochs. After these 150 epochs, a learning curve flattened. We did not experiment with different parameters of the Nadam optimiser and left them equal to their default values ($\beta_1 = 0.9$, $\beta_2 = 0.999$). A batch size equals 128. We denote this version SUPPNet (synth, S).

For active learning, the SUPPNet (synth) model was loaded and the learning continued for the next 100 epochs (for first 50 epochs $lr=10^{-4}$, and later $lr=10^{-5}$), using the same augmentations but with 10 percent of the synthetic training, examples replaced with spectra and \textit{pseudo-continua} acquired from manual normalisation. This resulted in SUPPNet (active, A) network.

All neural network outputs were used in the training process. As in previous experiments, a mean squared error was used as a loss for \textit{pseudo-continuum} regression, and mean binary cross-entropy for segmentation. The MSE loss was multiplied by a factor of 400 to give it the same order of magnitude as loss used in segmentation branches.

\subsection{Spectrum processing}

The proposed spectrum normalisation method consists of three stages. First, the spectrum is re-sampled with the sampling of 0.05\,\AA, and the sliding window of the size equal 8192 samples with the shift of 256 samples is applied. This particular shift means that each sample is normalised $32=8192/256$ times. The data prepared in this way is then re-scaled (min-max normalisation is applied). The second step is the application of the SUPP Network. As the post-processing step, the predicted \textit{pseudo-continua} are scaled back. Finally, the result is calculated as a weighted average over all predictions for each sample. This is also used to obtain a qualitative measure of the model's inherent uncertainty of both predictions (\textit{pseudo-continuum} and segmentation).

The processing described above gives final \textit{pseudo-continuum} and segmentation mask predictions. Because of the noisy pattern of the order of 0.001 of the relative magnitude present in the final result, we recommended an application of additional final post-processing. We used a smoothing spline available in SciPy Python module \citep[function \texttt{UnivariateSpline}]{2020SciPy-NMeth}. A useful advantage of this approach toward smoothing is the possibility of incorporating the estimated \textit{pseudo-continuum} errors in the smoothing process. A smoothing spline fit contains a knots arrangement that takes into account the uncertainty of the predicted \textit{pseudo-continuum}. For example, it places fewer knots in the ranges of greater uncertainty. Example SUPPNet result can be found in Fig.\,\ref{fig:example_result}.

\section{Results}

Several statistics were used for SUPPNet normalisation quality assessment. All were computed using residuals between the normalised spectrum (manually or automatically using SUPPNet) and the reference normalised spectrum. Either the synthetic or manually normalised spectrum served as the reference spectrum. Observed spectra analysed here come from UVES POP, while synthetic spectra were computed using ATLAS/SYNTHE codes. Among all normalised UVES POP spectra, six were chosen as representative and were normalised by three of us (TR, EN, and NP). These spectra show most features present in stellar spectra (for a detailed description see the following text and Table\,\ref{tab:stars_for_detailed_analysis}). Statistics mostly used in this section are the following percentiles: 2.28, 15.87, 50.00 (median), 84.13, and 97.73 and root-mean-square (RMS) error. In the case of normally distributed residuals 15.87--84.13 and 2.28--97.73 ranges would correspond to respectively one- and two-sigma bands.

The significance of the obtained results was tested using the bootstrap method. We tested the hypothesis that median and measure of spread, defined as a difference between 15.87 and 84.13 residuals, are equal in tested groups. In all statements regarding significance, we adopted 95\% symmetrical confidence intervals.

\begin{table*}
\caption{Stars for the detailed normalisation quality assessment.}
\label{tab:stars_for_detailed_analysis}
\centering
\begin{tabular}{lllllll}
\hline\hline
HD number & Name          & Spectral type & V {[}mag{]} & $T_\textrm{eff}$ {[}K{]} & $\log g$ & $v \sin i$  {[}km\,s$^{-1}${]}  \\ \hline
155806    & HR\,6397      & O7.5\,IIIe    & 5.53        &     -               & -        & 91~(1)    \\
90882     & $\delta$\,Sex & B9.5\,V       & 5.18        & 10139~(2)           & -        & 152~(3)   \\
27411     & HR\,1353      & A3m           & 6.06        & 7600~(4)            & 4.0~(4)  & 20.5~(4)  \\
37495     & $\nu^2$\,Col  & F5\,V         & 5.31        & 6417~(5)            & 3.79~(5) & 27.2~(6)  \\
59967     & HR\,2882      & G3\,V         & 6.64        & 5836~(7)            & 4.53~(7) & 3.76~(8)  \\
25069    & HR\,1232       &  K0\,III    & 5.83        & 4917~(9)            & 3.24~(9) & 3.24~(9)   \\ 
\hline
\end{tabular}
\tablebib{(1)~\citet{1997MNRAS.284..265H};
(2)~\citet{2012A&A...537A.120Z};
(3)~\citet{2017AJ....153...16S};
(4)~\citet{Catanzaro_2012};
(5)~\citet{Gomez_2018};
(6)~\citet{Schroder_2009};
(7)~\citet{Nissen_2020};
(8)~\citet{Santos_2016};
(9)~\citet{2019A&A...629A..80H}
}
\end{table*}

\subsection{Synthetic spectra normalisation}
\label{sec:Synthetic_spectra_normalisation}

To measure normalisation quality and minimise uncertainty introduced by the manual normalisation, the first test used only synthetic spectra. In the beginning, six chosen stars were modelled using ATLAS/SYNTHE codes. Parameters for synthetic spectra were taken from Table\,\ref{tab:stars_for_detailed_analysis} and related articles. Missing parameters were manually estimated to be equal $T_\textrm{eff}=30000$\,K, $\log g = 3.15$ in the case of HD\,155806 (O7.5\,IIIe) star, and $\log g = 3.90$ for HD\,90882 (B9.5\,V). Then, the synthetic normalised spectrum for each star was multiplied by about 200 different \textit{pseudo-continua} derived from the manual normalisation. That gave about 1200 spectra in total, which were later normalised using the two versions of SUPPNet (synth and active), and used to calculate normalisation metrics.

\begin{table*}
\caption{Summary statistics of the detailed analysis of chosen representative stars and the normalisation of synthetic spectra. Each value is reported in the following format: the main number is the median of residuals with 15.87 percentile in the upper index and 84.13 percentile in the lower index. The first two rows summarise the quality of normalisation of related synthetic spectra distorted with \textit{pseudo-continuum} fits obtained from manual normalisation (see Section \ref{sec:Synthetic_spectra_normalisation} for details). Here, contrary to the results on observed spectra, the correct normalisation is known, and the best values are in bold. Four bottom rows contain the statistics of SUPPNet and manual normalisation residuals. Manual normalisation of TR was used as the reference. All those residuals and statistics were calculated in the wavelength range from 3900\AA~to 7000\AA~(see the Section \ref{sec:detailed_analysis} for details).}

\label{tab:summary_statistics}
\centering
\renewcommand\arraystretch{1.3}
\begin{tabular}{lrrrrrr}
\hline\hline
Star & \multicolumn{1}{l}{HD\,155806} & \multicolumn{1}{l}{HD\,90882} & \multicolumn{1}{l}{HD\,27411} & \multicolumn{1}{l}{HD\,37495} & \multicolumn{1}{l}{HD\,59967} & \multicolumn{1}{l}{HD\,25069} \\ 
Spectral type &  \multicolumn{1}{l}{O7.5\,V} & \multicolumn{1}{l}{B9.5\,V} & \multicolumn{1}{l}{A3m} & \multicolumn{1}{l}{F4\,V} & \multicolumn{1}{l}{G4\,V} & \multicolumn{1}{l}{K0\,III} \\
\hline
SUPPNet (synth) &  $\mathbf{-0.0004^{0.0001}_{-0.0021}}$ & $0.0003^{0.0023}_{-0.0028}$ &  $0.0017^{0.0066}_{-0.0014}$ &  $0.0006\mathbf{^{0.0037}_{-0.0026}}$ &  $\mathbf{0.0005}^{0.0047}_{-0.0026}$ &  $\mathbf{0.0006}^{0.0057}_{-0.0038}$ \\
SUPPNet (active) &  $-0.0008^{-0.0002}_{-0.0027}$ & $-0.0003\mathbf{^{0.0016}_{-0.0049}}$ &  $\mathbf{0.0004^{0.0032}_{-0.0029}}$ &  $\mathbf{-0.0001}^{0.0021}_{-0.0043}$ &  $-0.0008\mathbf{^{0.0022}_{-0.0042}}$ &  $-0.0016\mathbf{^{0.0020}_{-0.0064}}$ \\
\hline
SUPPNet (synth)  &  $-0.0012^{0.0040}_{-0.0131}$ & $-0.0001^{0.0030}_{-0.0028}$ &  $-0.0010^{0.0033}_{-0.0102}$ &  $-0.0015^{0.0029}_{-0.0084}$ &    $0.0007^{0.0057}_{-0.0033}$ & $0.0084^{0.0251}_{-0.0000}$ \\
SUPPNet (active)  &  $-0.0026^{0.0023}_{-0.0126}$ & $-0.0013^{0.0011}_{-0.0044}$ &  $-0.0028^{0.0013}_{-0.0183}$ &  $-0.0024^{0.0012}_{-0.0092}$ &   $-0.0009^{0.0029}_{-0.0055}$ & $0.0048^{0.0177}_{-0.0044}$ \\
NP--TR &  $-0.0042^{0.0006}_{-0.0176}$ & $-0.0015^{0.0014}_{-0.0082}$ &  $-0.0013^{0.0032}_{-0.0080}$ &  $-0.0013^{0.0026}_{-0.0056}$ &  $-0.0042^{-0.0012}_{-0.0347}$ & $-0.0042^{0.0019}_{-0.0199}$ \\
EN--TR &   $0.0007^{0.0059}_{-0.0067}$ & $0.0006^{0.0028}_{-0.0012}$ &   $0.0007^{0.0058}_{-0.0064}$ &   $0.0004^{0.0039}_{-0.0041}$ &    $0.0016^{0.0052}_{-0.0016}$ & $0.0013^{0.0119}_{-0.0078}$ \\
\hline
\end{tabular}
\end{table*}

The summary statistics of this experiment can be found in the top two rows of Table \ref{tab:summary_statistics}, in Fig.\,\ref{fig:synthetic_result_active} and Fig.\,\ref{fig:synthetic_result_synth} in Appendix \ref{appendix:graphs}. Medians of residuals, which measure a normalisation bias, are between $-0.0016$ (SUPPNet active, HD\,25069, K0\,III) and 0.0017 (SUPPNet synth, HD\,27411, A3m). Residuals in the case of SUPPNet trained using active learning are systematically smaller than when trained using only synthetic dataset. This means that the latter places \textit{pseudo-continuum} systematically lower. The dispersion of residuals measured as a difference between 84.13 and 15.87 percentiles are between 0.0022 (SUPPNet synth, HD\,155806, O7.5\,IIIe) and 0.0095 (SUPPNet synth, HD\,27411, A3m). The residuals are often slightly smaller when active learning is applied. For SUPPNet trained using the active learning the dispersion of residuals is slightly but significantly smaller in the case of HD\,27411 (A3m), HD\,37495 (G4\,V) and HD\,25069 (K0\,III), is significantly larger for HD\,90882 (B9.5\,V). There are no statistically significant differences between dispersions of residuals for HD\,155806 (O7.5\,V) and HD\,37495 (F4\,V). The detailed inspection of Fig.\,\ref{fig:synthetic_result_active} shows that systematic errors vary significantly across both wavelength and spectra parameters. They are especially prominent for A3m (HD\,27411) and F4V (HD\,37495) stars, in wavelength shorter than 4500\,\AA, where they reach 0.03. Figure\,\ref{fig:hd27411_A3m_synthetic_error} is a close-up of the 3900--4500\,\AA~wavelength range of the A3\,V synthetic spectra with the mean normalisation result and residuals. It can be seen, that a significant normalisation bias arises in a range where the whole spectrum is below the continuum level. Average biases in other wavelength ranges of synthetic spectra's residuals are generally below 0.01.

We consider this statistics to be close to the realistic normalisation uncertainty in wavelength ranges with spectral features well represented in synthetic models. Nonetheless, normalisation of the synthetic spectra does not allow us to draw conclusions about the normalisation quality of the observed spectra, as they contain additional features (e.g.  complex emission spectral lines).

\begin{figure*}
\resizebox{\hsize}{!}
{\includegraphics[clip]{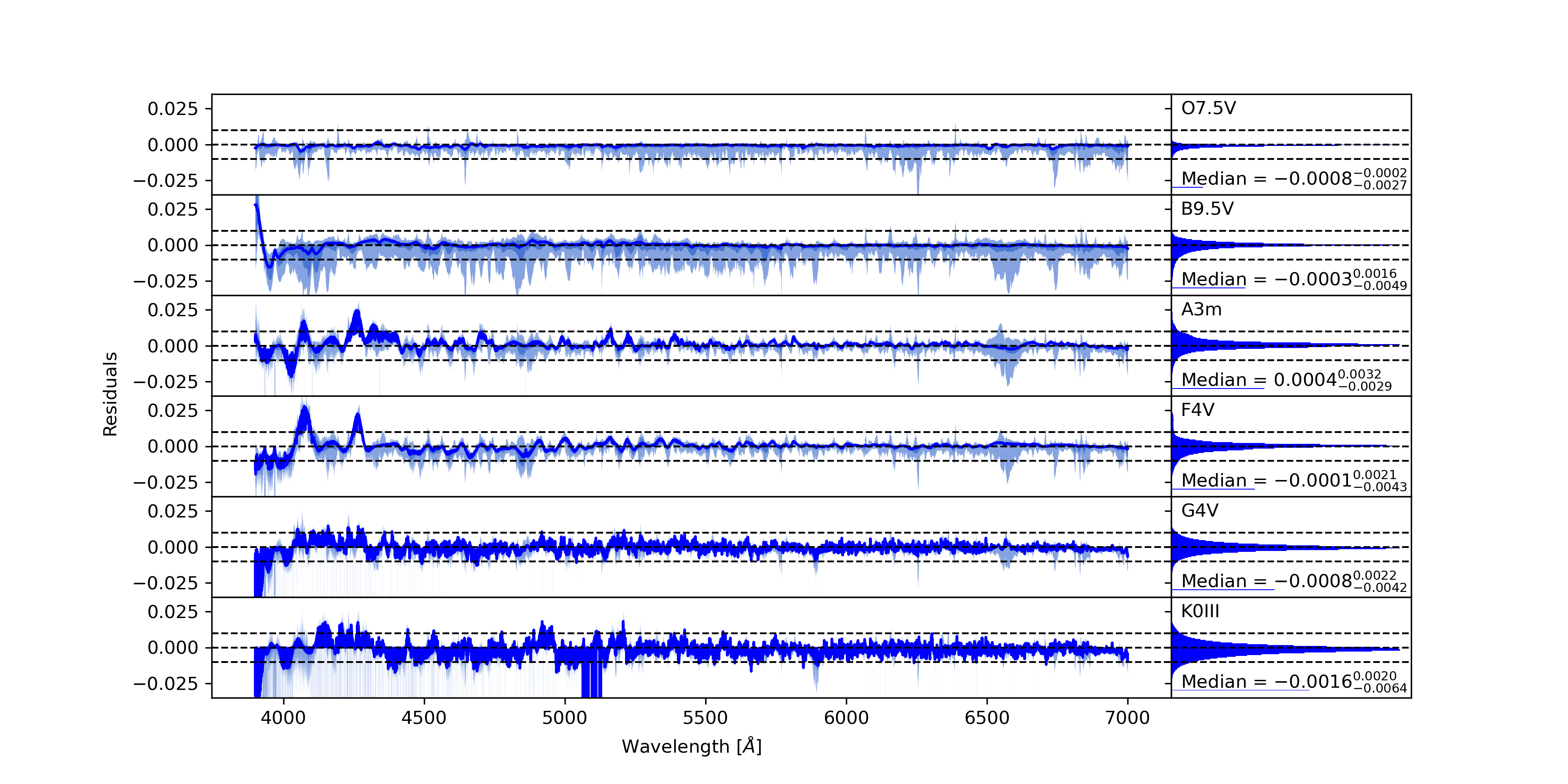}}
\caption{Results of normalisation of six synthetic spectra multiplied by six manually fitted \textit{pseudo-continua} trained with the application of \textbf{active learning} (synthetic data supplemented with manually normalised spectra). In each row, on the left, the differences between automatically normalised spectra and synthetic spectra are shown, and on the right, the histograms of those differences with a related spectral type, median with 15.87 percentile in the upper index, and 84.13 percentile in the lower index are displayed. The dashed lines on each panel correspond to the residuals equal -0.01, 0.0 and 0.01 respectively. The use of active learning resulted in a slight reduction of residuals' dispersion.}
\label{fig:synthetic_result_active}
\end{figure*}

\begin{figure}
\resizebox{\hsize}{!}
{\includegraphics[clip]{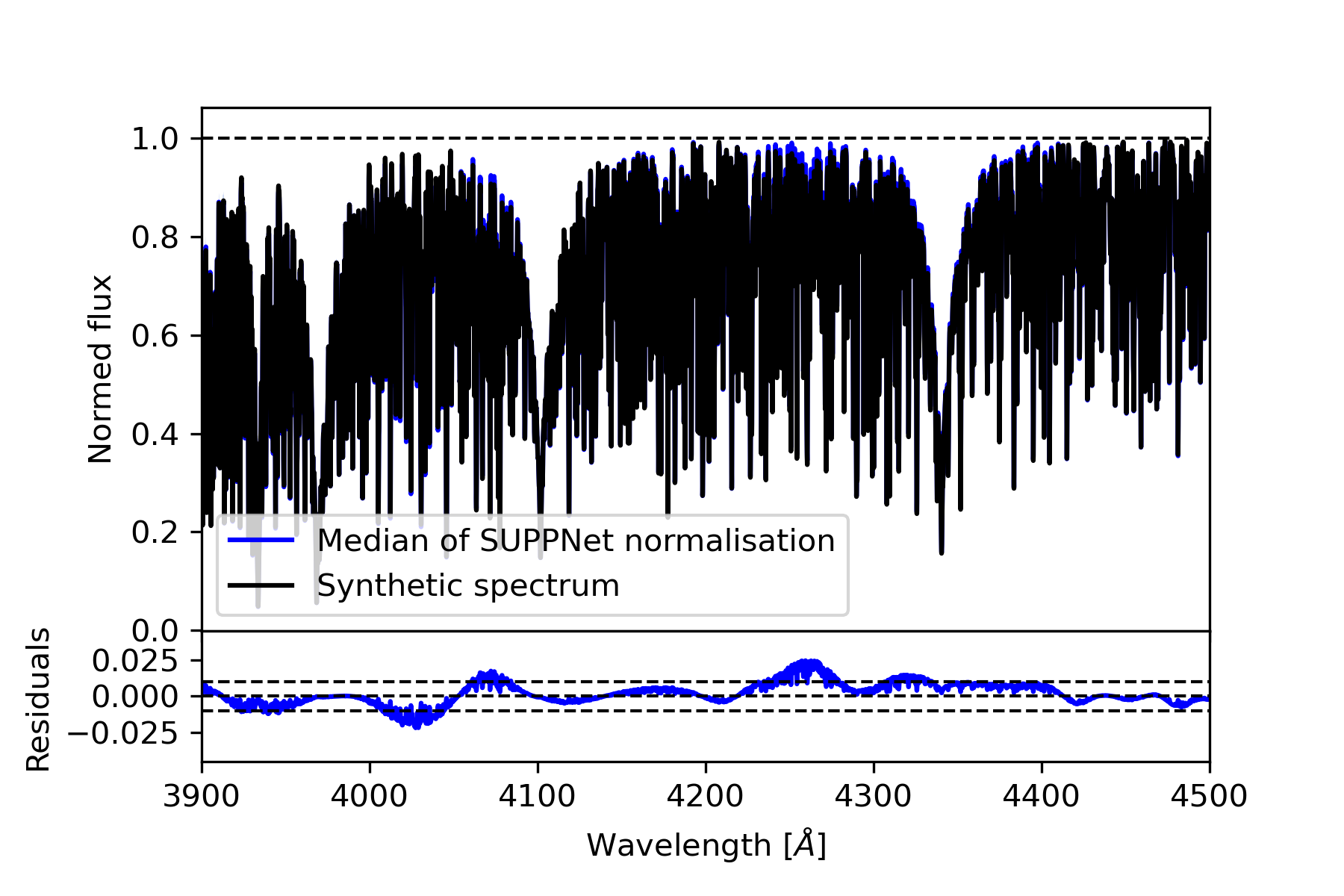}}
\caption{Close-up of the 3900--4500\,\AA~spectral range of the A3\,V synthetic spectrum with a median of synthetic automatically normalised spectra (top panel) and residuals of normalisation errors (bottom panel). For this particular part of the spectrum, the average normalisation is significantly biased. These differences arise due to wide hydrogen absorption lines and strong metal lines which heavily blend in this spectral range. See Fig.\,\ref{fig:synthetic_result_active} for the remaining results.}
\label{fig:hd27411_A3m_synthetic_error}
\end{figure}

\subsection{Summary UVES POP normalisation statistics}

\begin{figure*}
\resizebox{\hsize}{!}
{\includegraphics[clip]{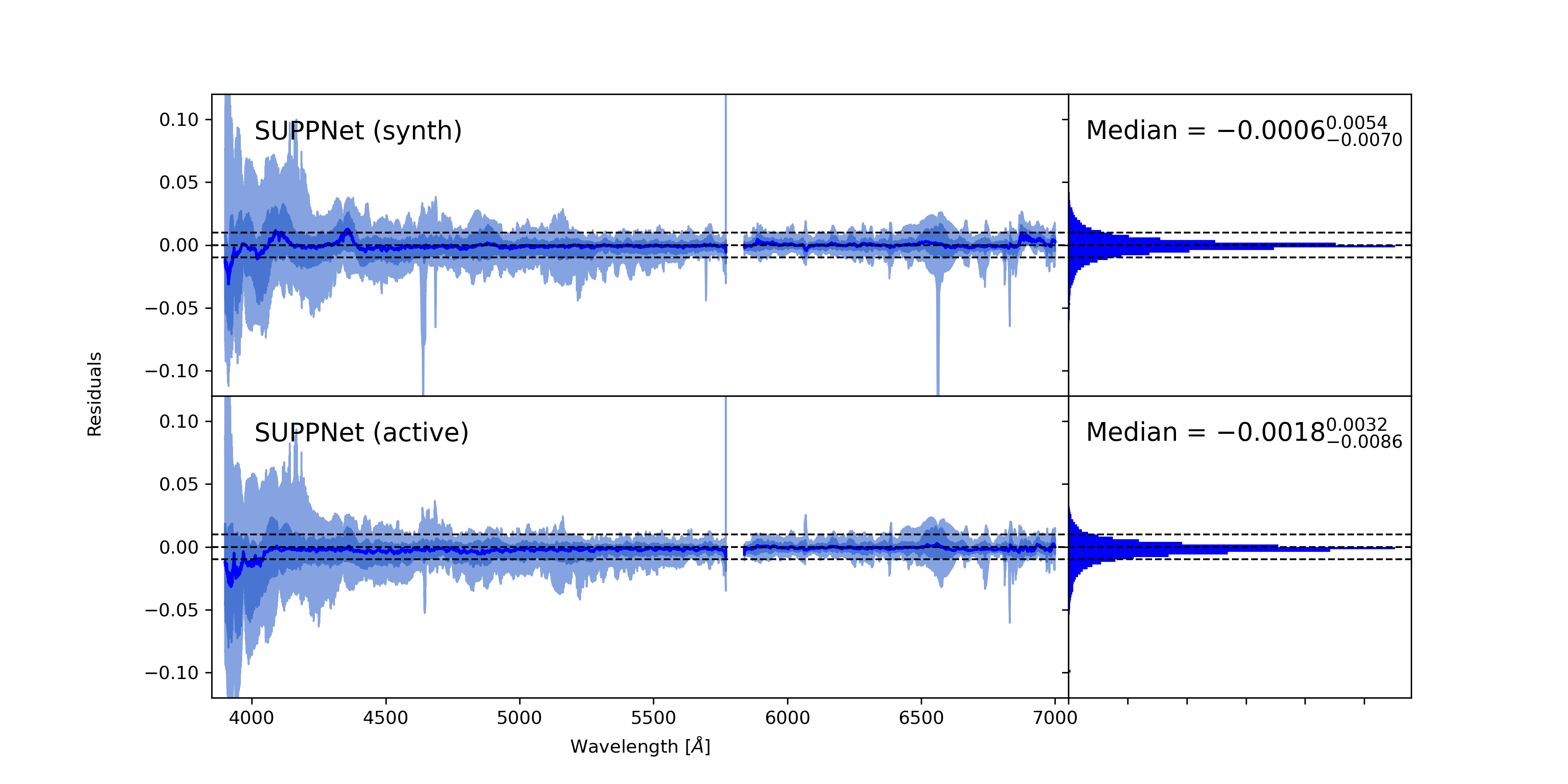}}
\caption{Quality of normalisation measured using residuals between the result of SUPPNet method and the manually normalised spectra over all stars from UVES POP field stars, that were manually normalised. The line shows the value of median for each wavelength, the shaded areas are defined to contain respectively 68 and 95 percent of values (defined by percentiles: 2.28, 15.87, 84.13, and 97.73). The upper panel contains the results of the algorithm that used only synthetic data for training, the lower one - that applied active learning. Active learning significantly reduces systematic effects for wavelengths shorter than 4500\,\AA.}
\label{fig:UVES_POP_percentiles}
\end{figure*}

Approximately a hundred spectra from the UVES POP library were used to assess the quality of SUPPNet on observed stellar spectra. In the beginning, the majority of spectra from spectral type O to G were manually normalised. Then spectra were normalised using both versions of SUPPNet, and residuals' statistics were calculated. The results are summarised in Fig.\,\ref{fig:UVES_POP_percentiles}.

In the case of the model trained using only synthetic data some biases are prominent around 6750\,\AA  and for wavelengths shorter than 4500\,\AA. Strong telluric lines at approximately 6870\,\AA, which were absent in synthetic spectra, are the cause of the first bias. The second problem appears because SUPPNet (synth) systematically places \textit{pseudo-continuum} too low in wide hydrogen absorption lines of A and F spectra types. Check Figs.\,\ref{fig:UVES_POP_percentiles_hist_typeO}-\ref{fig:UVES_POP_percentiles_hist_typeG} in Appendix\,\ref{appendix:graphs} for separate plots for each spectral type. The median of residuals is significantly different from 0 and equals $-0.0006$, which means that the model places \textit{pseudo-continuum} slightly above levels chosen by astronomers. The dispersion of residuals measured as a difference between 84.13 and 15.87 percentiles equals 0.0124. Root mean squared error equals 0.0122.

SUPPNet trained using an active learning approach shows slightly better characteristics. Dispersion of its residuals (0.0118, $\textrm{RMS}=0.0128$) is not significantly different from the SUPPNet (synth), but most of the systematic effects that can be seen in Fig.\,\ref{fig:UVES_POP_percentiles} are reduced (compare bias in wavelengths shorter than 4500\,\AA) after training that included real spectra. Nonetheless, SUPPNet (active) shares the tendency of the model trained on a synthetic dataset, to place the \textit{pseudo-continuum} higher than when a spectrum is manually normalised. This tendency is measured by a median of residuals which equals $-0.0018$ and is significantly (for a 95\% confidence interval) different from 0. This effect can potentially be introduced by the human, who often model \textit{pseudo-continuum} to lie lower than in reality. Differences in medians of residuals for both SUPPNet versions, although statistically significant, are at most comparable and often smaller than an intrinsic error of manual normalisation described in detail in Section \ref{sec:detailed_analysis}.

Normalisation quality and active learning importance vary significantly across spectral types. In general the later the spectral type is the larger is the dispersion of normalisation residuals. Differences between residuals' medians of both versions of SUPPNets for O, B, and A type stars are statistically significant. However, SUPPNet synth and active are not significantly different in terms of residuals' dispersions and remaining medians. The results are summarised in Table\,\ref{tab:UVES_POP_spectral_types_summary} and summary plots for each spectral type can be found in Figs.\,\ref{fig:UVES_POP_percentiles_hist_typeO}-\ref{fig:UVES_POP_percentiles_hist_typeG} in Appendix\,\ref{appendix:graphs}. The tendency to place \textit{pseudo-continuum} higher than during manual normalisation holds for most spectral types, with a notable exception of G type stars. Inspection showed that in spectral ranges with substantial narrow absorption lines blending, where the continuum is absent across wide parts of a spectrum, the model tends to place \textit{pseudo-continuum} below the correct level. A simple workaround for this SUPPNet limitation is to reduce default sampling from 0.05\,\AA~to 0.03--0.04\,\AA, when working with spectral types later than G5.

\begin{table}[]
\caption{Summary of SUPPNet normalisation quality over UVES POP field stars divided into spectral types. The model generally predicts \textit{pseudo-continuum} higher than a human. The prominent except here are spectra of G type stars where this tendency is inverted. The detailed plots can be found in Figs.\,\ref{fig:UVES_POP_percentiles_hist_typeO}-\ref{fig:UVES_POP_percentiles_hist_typeG} in Appendix\,\ref{appendix:graphs}.}
\label{tab:UVES_POP_spectral_types_summary}
\centering
\begin{tabular}{crr}
\hline\hline
\multicolumn{1}{l}{Spectral type} & \multicolumn{1}{l}{SUPPNet (synth)} & \multicolumn{1}{l}{SUPPNet (active)}\\ \hline
O & $-0.0011^{0.0010}_{-0.0041}$ & $-0.0025^{-0.0004}_{-0.0056}$\\
B & $-0.0003^{0.0025}_{-0.0039}$ & $-0.0014^{0.0010}_{-0.0053}$\\
A &  $0.0005^{0.0079}_{-0.0060}$ & $-0.0006^{0.0054}_{-0.0076}$\\
F & $-0.0031^{0.0026}_{-0.0109}$ & $-0.0042^{0.0007}_{-0.0129}$\\
G & $0.0021^{0.0107}_{-0.0045}$ & $0.0005^{0.0079}_{-0.0061}$\\
\hline
\end{tabular}
\end{table}

\begin{figure}
\resizebox{\hsize}{!}
{\includegraphics[clip]{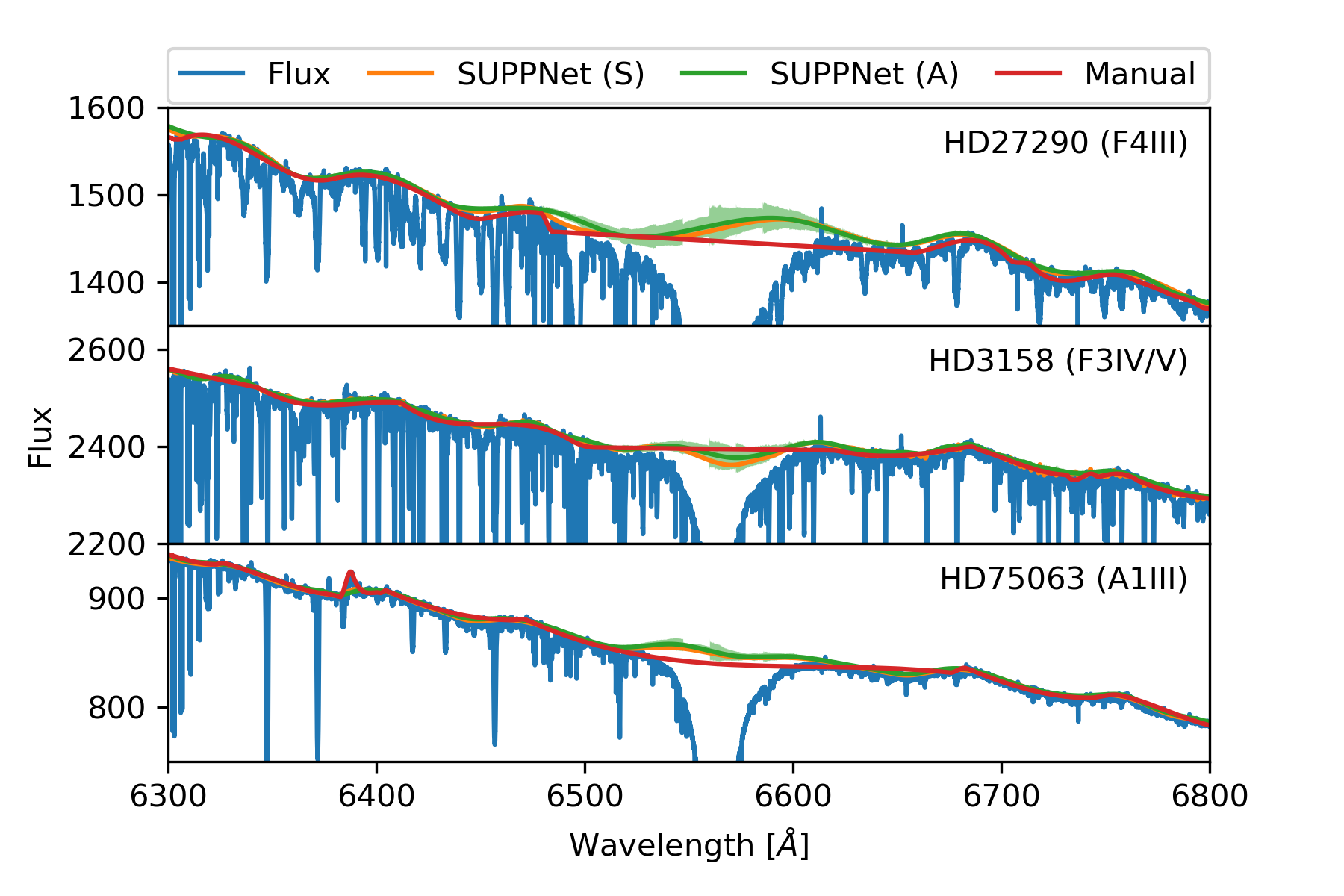}}
\caption{H$\alpha$ Balmer line region for the three UVES POP field stars. The figure shows how a wavy pattern prominent in \textit{pseudo-continua} of F and A type stars is related to manual and SUPPNet predictions. The \textit{pseudo-continuum} predicted by SUPPNet (A) is shown with an estimate of its uncertainty (method internal uncertainty, green shaded area).}
\label{fig:Halpha_zoom}
\end{figure}

The second issue is the relatively high uncertainty in the H$\alpha$ Balmer line, prominent especially in the case of F  and A type stars. Closer examination showed that this can be explained by erroneous manual normalisation. For this spectral range, the manual normalisation is impossible as a wavy pattern introduced by imperfections of an instrument pipeline crosses and changes this spectral feature significantly (see Fig.\,\ref{fig:Halpha_zoom} for details).

\begin{figure}
\resizebox{\hsize}{!}
{\includegraphics[clip]{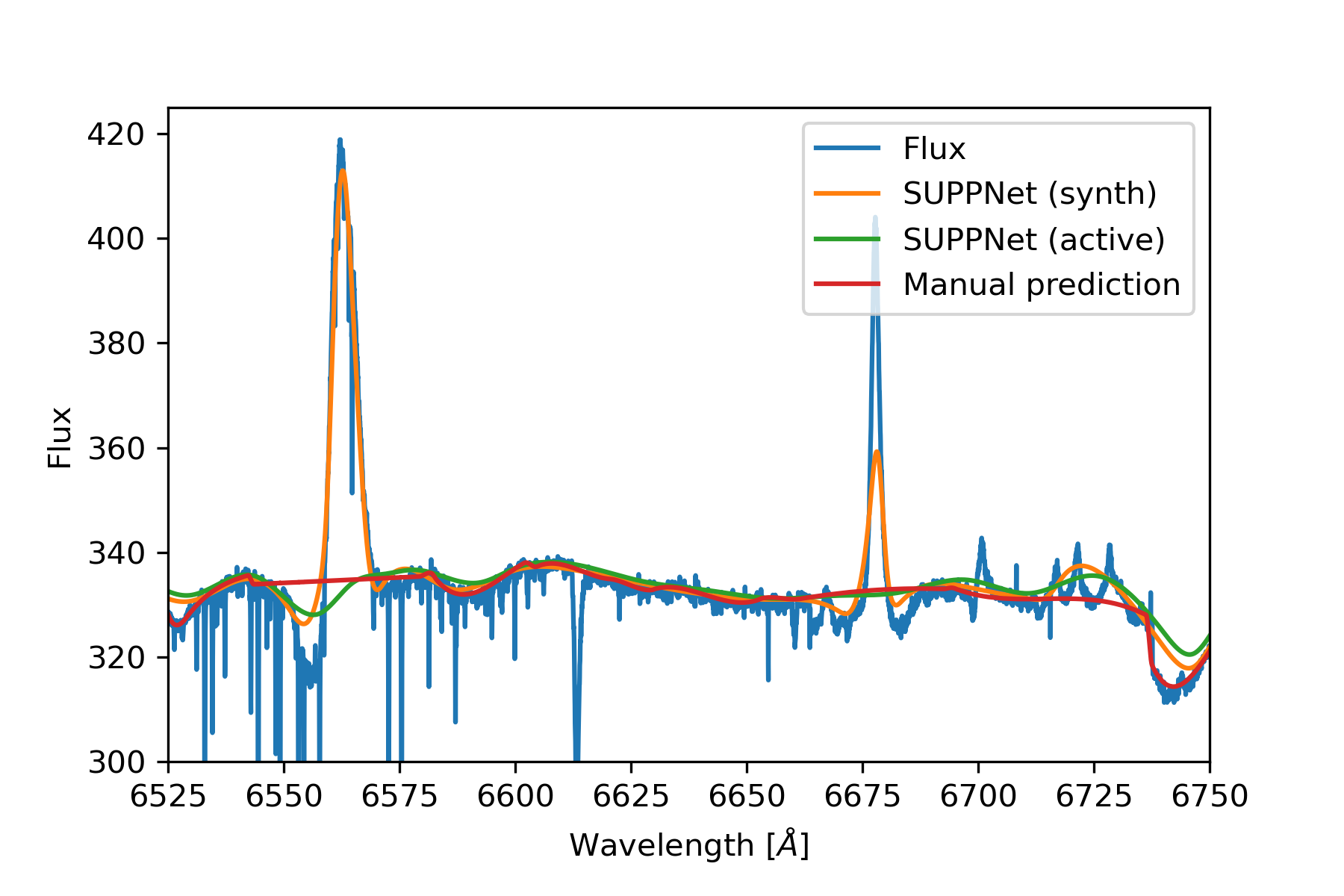}}
\caption{Predicted \textit{pseudo-continuum} for a spectrum of HD\,148937\,(O6.5) with H$\alpha$ and HeI\,6678\,\AA~ lines in emission. SUPPNet (active) correctly deals with most emission features, while SUPPNet (synth) treats those features as a part of \textit{pseudo-continuum}. This is an important example where active learning significantly improves the normalisation quality.}
\label{fig:example_emmision_lines}
\end{figure}

The positive influence of active learning on SUPPNet normalisation quality is the most prominent in the case of O type stars, where strong emission features are often present. A spectral range of the HD\,148937\,(O6.5) star, with H$\alpha$ Balmer and He\,I\,6678\,\AA~lines in emission is shown in the Fig.\,\ref{fig:example_emmision_lines} . SUPPNet (active) relatively well predicts \textit{pseudo-continuum} across these features while SUPPNet, trained using synthetic data only, treats these features as a part of \textit{pseudo-continuum}.

\subsection{Detailed analysis of chosen stars}
\label{sec:detailed_analysis}

In the last step of the SUPPNet analysis, six stars were normalised automatically and manually by three of us (TR, EN and NP), and carefully compared. Their spectral types range from O7.5\,V to K0\,III and show most of the typical spectral features. The main characteristics of the selected stars are gathered in Table\,\ref{tab:stars_for_detailed_analysis}.

HD\,155806 (HR\,6397, $V = 5.53$~mag) is the hottest Galactic Oe star \citep{2011IAUS..272..182F}. Its initial spectral classification O7.5\,V[n]e \citep{1973AJ.....78.1067W} was changed to O7.5\,IIIe because of the strength of its metallic features \citep{2004AN....325..749N}. Stars of this type are rare and show a double-peaked or central emission in their Balmer lines.

HD\,27411 (HR\,1353, $V = 6.06$~mag, A3m) is an example of a chemically peculiar (CP) star. Its atmospheric parameters and abundances were investigated in detail in the context of diffusion theory in the work of \citet{Catanzaro_2012}.

HD\,90882 (HR\,4116, $\delta$\,Sex, B9.5\,V), HD\,37495 (HR\,1935, $\nu^2$\,Col, F5\,V), and HD\,59967 (HR\,2882, G3\,V) are typical representatives of their spectral types. The first is a rapidly rotating B type star, and the last is a young ($\approx0.4$\,Gyr), active, slowly rotating solar-twin star.

The last star for detailed analysis is HD\,25069 (HR\,1232, $V=5.83$~mag). In the UVES POP catalogue, its spectral type is G9\,V, while SIMBAD's database sources give K0\,III or K0/1\,III. Here K0\,III is used. This star is a representative example of a late G and early K spectral type.

\begin{figure*}
\resizebox{\hsize}{!}
{\includegraphics[clip]{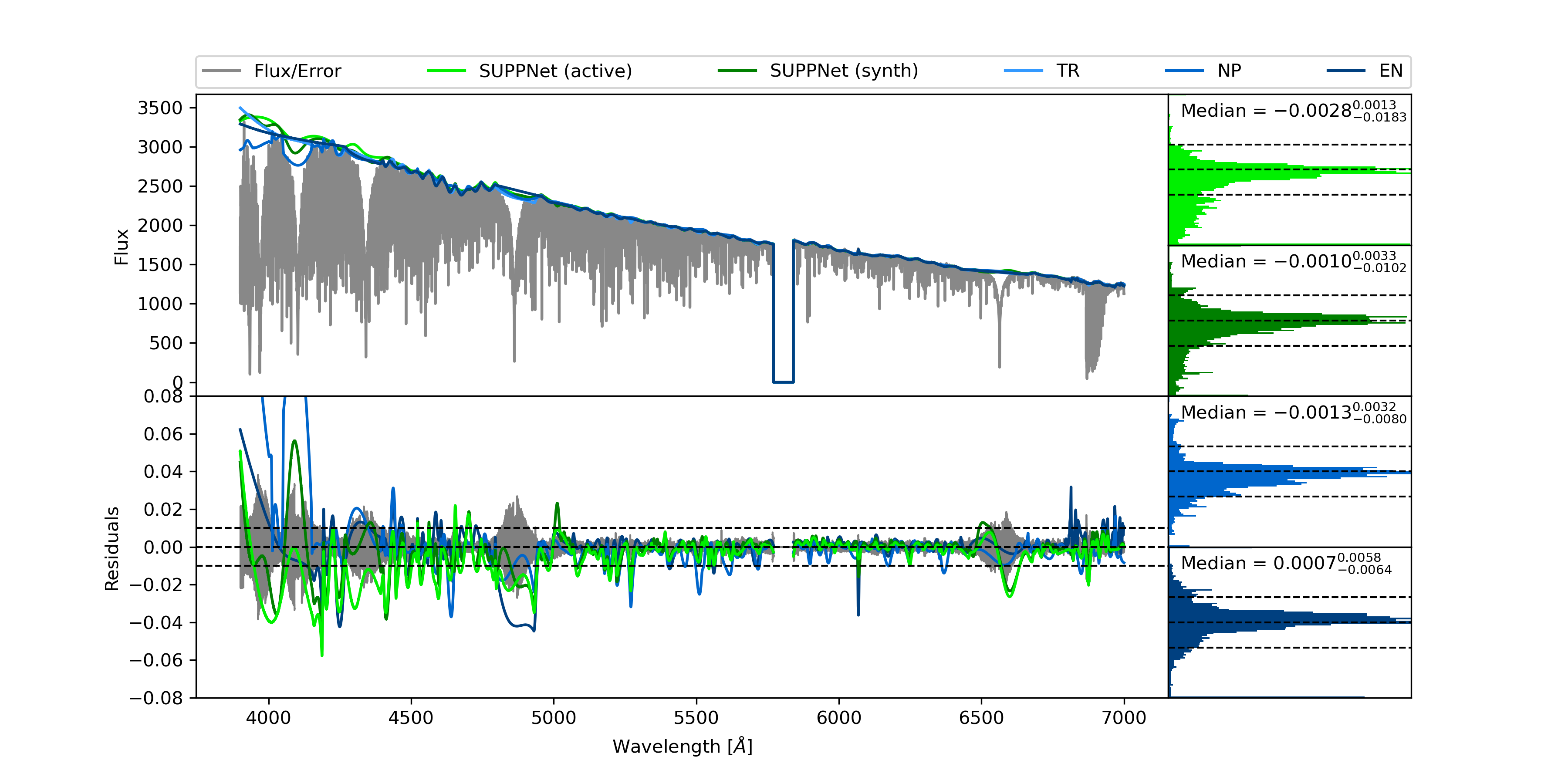}}
\caption{Comparison of normalisation quality on the example of HD\,27411 (A3m) star with two versions of the proposed method (SUPPNet active and synth) and manual normalisation done independently by three different people (TR, NP, and EN). The left upper panel shows original flux with all fitted \textit{pseudo-continua}. The left panel shows residuals of normalised fluxes relative to TR normalisation. The right panel presents histograms of all mentioned residuals with median, 15.87, and 84.13 percentiles.}
\label{fig:27411_manual}
\end{figure*}

The results for all stars can be found in the four bottom rows of Table \ref{tab:summary_statistics}. Figure\,\ref{fig:27411_manual} contains detailed \textit{pseudo-continua} and residuals for the HD\,27411 (A3m) star. The results show that the differences between manually normalised spectra are between $-0.0042$ and $0.0016$ in the median with typical dispersion, defined by a 15.87--84.13 percentiles band, ranging approximately from 0.0040 to 0.0330 (see NP--TR and EN--TR rows in Table\,\ref{tab:summary_statistics}). This is the scale of uncertainty inherent in manual normalisation. In terms of residual's statistics, the quality of the SUPPNet normalisation method is superior in comparison to the quality of the manual normalisation.

The left panels of Fig.\,\ref{fig:27411_manual} and Figs.\,\ref{fig:155806_manual}-\ref{fig:25069_manual} in Appendix \ref{appendix:graphs} convince why the \textit{pseudo-continuum} fitting is a challenging task. In HD\,27411 most of the spectrum is disturbed by a semi-periodic pattern that arises from imperfect orders merging and a blaze function removal. SUPPNet models this \textit{pseudo-continuum} type relatively well for wavelengths longer than the Balmer H$\beta$ line and significantly worse in a spectral range from 4400 to 4800\,\AA, where the amplitude and frequency of this pattern significantly increase. Nonetheless, the error amplitude is of the order of 0.02, both for manual and SUPPNet normalisation. For wavelengths shorter than 4200\,\AA\, the dispersion between different normalised fluxes grows considerably. In this range, it is difficult to assess the normalisation quality without referring to the synthetic spectral model, which could potentially guide manual normalisation. 

The uncertainty estimated by SUPPNets has only qualitative meaning and informs the users where they can expect normalisation results to be most uncertain, but not necessarily where the model failed in predicting \textit{pseudo-continuum}. As can be seen in the bottom panel of Fig.\,\ref{fig:27411_manual} the estimated uncertainty is the highest in the wide absorption lines. However, in the spectral range from 4400 to 4800\,\AA~the proposed method did not capture the ambiguous character of the predicted \textit{pseudo-continuum}. 

\subsection{Resolution, rotational velocity, and noise}

Stellar spectra present in databases have various levels of noise, resolutions, and projected rotational velocities. We tested the consistency of normalisation with respect to those parameters.
For that purpose synthetic spectra calculated for parameters of HD\,27411\,(A3m) with a signal-to-noise equal 30 and 500, a projected rotational velocity equal 0, 50, 100, 200\,km~s$^{-1}$, and resolution equal $10^4$ and $10^5$ were used.

Predicted \textit{pseudo-continuum} is generally consistent for the tested signal-to-noise ratios. The differences between results increase for low-resolution spectra and in heavily blended line regions where there is no real continuum in the flux. The test with low-quality medium-resolution spectra showed that in this case, the proposed method places the \textit{pseudo-continuum} significantly too high. This problem can be suppressed by increasing a sampling step or by training SUPPNet using low-resolution, noisy spectra.

For a spectrum broadened due to the projected rotational velocity, the differences between predicted \textit{pseudo-continua} are generally smaller than 0.01, except for the wavelengths shorter than 4300\,\AA~where a flux is well below expected \textit{pseudo-continuum}.

The results described above are summarised in Fig.\,\ref{fig:resolution_and_vsini_influence} in Appendix\,\ref{appendix:noise}.

\section{Conclusions}

Machine learning methods are becoming more and more important in science, particularly in astronomy. Here we had presented a method for spectrum normalisation that uses novel, one dimensional, fully convolutional, deep neural network architecture -- SUPP Network. Its automatic character makes SUPPNet's results reproducible and more homogeneous which is very important especially in the context of time series of spectra. Usage of synthetic spectra during the training makes the model aware of features present in spectra and helps to recover \textit{pseudo-continuum} in regions where it cannot be done manually, e.g. in regions heavily blended by lines or across Balmer hydrogen lines with instrumental ripples. The accuracy of SUPPNet places it next to careful manual normalisation and makes it possible to omit human intervention in this step of spectrum pre-processing. It works well both with emission and absorption spectral features, with blended lines and spectral ranges where the real continuum is absent. If manual correction is necessary it can be done using a developed Python application or online, see Appendix\,\ref{appendix:codes}.

The main drawback of the proposed method is the fact that it uses a sliding window technique while in principle it is not necessary for fully convolutional neural networks, as they accept inputs of any length. Sliding window is necessary, as we used a min-max normalisation strategy for inputs. As a beneficial side effect, we are able to give some estimates of \textit{pseudo-continuum} uncertainty.

Directions of future developments are the following: extension of the training set to include spectra from spectrographs other than UVES and FEROS, an extension of the set of empirical \textit{pseudo-continua} to include fits from other astronomers, experiments with alternative input normalisation techniques that would eliminate the sliding window approach, and exploration of alternative ways to estimate prediction uncertainties. Other astronomers' \textit{pseudo-continua} and observations from different instruments are expected to reduce biases. The proposed technique can be extended in such a way that it uses coarse spectral type estimation and wavelength range information, which can further improve normalisation quality. Other possible development paths are exploitation of different neural network architectures and modules (e.g. attention module) and scaling the SUPPNet model up. The longstanding purpose is to use an automatically normalised spectrum for stellar parameters' and abundances' estimation that consistently includes normalisation errors.

The most important characteristic of the proposed method is its generality. It is straightforward to use the SUPP Network in similar tasks, like a trend or background modelling, and also use UPP modules in contexts other than related to one-dimensional signals processing, e.g. in image segmentation.

\begin{acknowledgements}
T. R\'o\.za\'nski was financed from budgetary funds for science in 2018-2022 as a research project under the program "Diamentowy Grant", no. DI2018 024648. 
Research project partly supported by program "Excellence initiative - research university" for years 2020-2026 for University of Wroc{\l}aw.
\end{acknowledgements}

%


\begin{thebibliography}{54}
\expandafter\ifx\csname natexlab\endcsname\relax\def\natexlab#1{#1}\fi

\bibitem[{Abadi {et~al.}(2015)Abadi, Agarwal, Barham, Brevdo, Chen, Citro,
  Corrado, Davis, Dean, Devin, Ghemawat, Goodfellow, Harp, Irving, Isard, Jia,
  Jozefowicz, Kaiser, Kudlur, Levenberg, Man\'{e}, Monga, Moore, Murray, Olah,
  Schuster, Shlens, Steiner, Sutskever, Talwar, Tucker, Vanhoucke, Vasudevan,
  Vi\'{e}gas, Vinyals, Warden, Wattenberg, Wicke, Yu, \&
  Zheng}]{tensorflow2015-whitepaper}
Abadi, M., Agarwal, A., Barham, P., {et~al.} 2015, {TensorFlow}: Large-Scale
  Machine Learning on Heterogeneous Systems, software available from
  tensorflow.org

\bibitem[{{Aguilera-G{\'o}mez} {et~al.}(2018){Aguilera-G{\'o}mez},
  {Ram{\'\i}rez}, \& {Chanam{\'e}}}]{Gomez_2018}
{Aguilera-G{\'o}mez}, C., {Ram{\'\i}rez}, I., \& {Chanam{\'e}}, J. 2018, \aap,
  614, A55

\bibitem[{{Antoniadis-Karnavas, A.} {et~al.}(2020){Antoniadis-Karnavas, A.},
  {Sousa, S. G.}, {Delgado-Mena, E.}, {Santos, N. C.}, {Teixeira, G. D. C.}, \&
  {Neves, V.}}]{antoniadis2020}
{Antoniadis-Karnavas, A.}, {Sousa, S. G.}, {Delgado-Mena, E.}, {et~al.} 2020,
  A\&A, 636, A9

\bibitem[{Bagnulo {et~al.}(2003)Bagnulo, Jehin, Ledoux, Cabanac, Melo,
  Gilmozzi, {et~al.}}]{bagnulo2003uves}
Bagnulo, S., Jehin, E., Ledoux, C., {et~al.} 2003, Messenger, 114, 10

\bibitem[{Ball \& Brunner(2010)}]{BALL_2010}
Ball, N.~M. \& Brunner, R.~J. 2010, International Journal of Modern Physics D,
  19, 1049–1106

\bibitem[{Baron(2019)}]{baron2019machine}
Baron, D. 2019, Machine Learning in Astronomy: a practical overview

\bibitem[{Cadusch {et~al.}(2013)Cadusch, Hlaing, Wade, McArthur, \&
  Stoddart}]{Cadusch_2013}
Cadusch, P.~J., Hlaing, M.~M., Wade, S.~A., McArthur, S.~L., \& Stoddart, P.~R.
  2013, Journal of Raman Spectroscopy, 44, 1587–1595

\bibitem[{Carleo {et~al.}(2019)Carleo, Cirac, Cranmer, Daudet, Schuld, Tishby,
  Vogt-Maranto, \& Zdeborová}]{Carleo_2019}
Carleo, G., Cirac, I., Cranmer, K., {et~al.} 2019, Reviews of Modern Physics,
  91

\bibitem[{Catanzaro \& Balona(2012)}]{Catanzaro_2012}
Catanzaro, G. \& Balona, L.~A. 2012, Monthly Notices of the Royal Astronomical
  Society, 421, 1222

\bibitem[{Cretignier {et~al.}(2020)Cretignier, Francfort, Dumusque, Allart, \&
  Pepe}]{Cretignier_2020}
Cretignier, M., Francfort, J., Dumusque, X., Allart, R., \& Pepe, F. 2020,
  Astronomy \& Astrophysics, 640, A42

\bibitem[{{dos Santos} {et~al.}(2016){dos Santos}, {Mel{\'e}ndez}, {do
  Nascimento}, {Bedell}, {Ram{\'\i}rez}, {Bean}, {Asplund}, {Spina},
  {Dreizler}, {Alves-Brito}, \& {Casagrande}}]{Santos_2016}
{dos Santos}, L.~A., {Mel{\'e}ndez}, J., {do Nascimento}, J.-D., {et~al.} 2016,
  \aap, 592, A156

\bibitem[{Dozat(2016)}]{dozat2016incorporating}
Dozat, T. 2016

\bibitem[{Farias {et~al.}(2020)Farias, Ortiz, Damke, {Jaque Arancibia}, \&
  Solar}]{FARIAS2020100420}
Farias, H., Ortiz, D., Damke, G., {Jaque Arancibia}, M., \& Solar, M. 2020,
  Astronomy and Computing, 33, 100420

\bibitem[{{Fullerton} {et~al.}(2011){Fullerton}, {Petit}, {Bagnulo}, {Wade}, \&
  {Wade}}]{2011IAUS..272..182F}
{Fullerton}, A.~W., {Petit}, V., {Bagnulo}, S., {Wade}, G.~A., \& {Wade}. 2011,
  in Active OB Stars: Structure, Evolution, Mass Loss, and Critical Limits, ed.
  C.~{Neiner}, G.~{Wade}, G.~{Meynet}, \& G.~{Peters}, Vol. 272, 182--183

\bibitem[{{Gaia Collaboration} {et~al.}(2016){Gaia Collaboration}, {Prusti},
  {de Bruijne}, {Brown}, {Vallenari}, {Babusiaux}, {Bailer-Jones}, {Bastian},
  {Biermann}, {Evans}, {Eyer}, {Jansen}, {Jordi}, {Klioner}, {Lammers},
  {Lindegren}, {Luri}, {Mignard}, {Milligan}, {Panem}, {Poinsignon},
  {Pourbaix}, {Randich}, {Sarri}, {Sartoretti}, {Siddiqui}, {Soubiran},
  {Valette}, {van Leeuwen}, {Walton}, {Aerts}, {Arenou}, {Cropper}, {Drimmel},
  {H{\o}g}, {Katz}, {Lattanzi}, {O'Mullane}, {Grebel}, {Holland}, {Huc},
  {Passot}, {Bramante}, {Cacciari}, {Casta{\~n}eda}, {Chaoul}, {Cheek}, {De
  Angeli}, {Fabricius}, {Guerra}, {Hern{\'a}ndez}, {Jean-Antoine-Piccolo},
  {Masana}, {Messineo}, {Mowlavi}, {Nienartowicz}, {Ord{\'o}{\~n}ez-Blanco},
  {Panuzzo}, {Portell}, {Richards}, {Riello}, {Seabroke}, {Tanga},
  {Th{\'e}venin}, {Torra}, {Els}, {Gracia-Abril}, {Comoretto},
  {Garcia-Reinaldos}, {Lock}, {Mercier}, {Altmann}, {Andrae}, {Astraatmadja},
  {Bellas-Velidis}, {Benson}, {Berthier}, {Blomme}, {Busso}, {Carry},
  {Cellino}, {Clementini}, {Cowell}, {Creevey}, {Cuypers}, {Davidson}, {De
  Ridder}, {de Torres}, {Delchambre}, {Dell'Oro}, {Ducourant}, {Fr{\'e}mat},
  {Garc{\'\i}a-Torres}, {Gosset}, {Halbwachs}, {Hambly}, {Harrison}, {Hauser},
  {Hestroffer}, {Hodgkin}, {Huckle}, {Hutton}, {Jasniewicz}, {Jordan},
  {Kontizas}, {Korn}, {Lanzafame}, {Manteiga}, {Moitinho}, {Muinonen},
  {Osinde}, {Pancino}, {Pauwels}, {Petit}, {Recio-Blanco}, {Robin}, {Sarro},
  {Siopis}, {Smith}, {Smith}, {Sozzetti}, {Thuillot}, {van Reeven}, {Viala},
  {Abbas}, {Abreu Aramburu}, {Accart}, {Aguado}, {Allan}, {Allasia},
  {Altavilla}, {{\'A}lvarez}, {Alves}, {Anderson}, {Andrei}, {Anglada Varela},
  {Antiche}, {Antoja}, {Ant{\'o}n}, {Arcay}, {Atzei}, {Ayache}, {Bach},
  {Baker}, {Balaguer-N{\'u}{\~n}ez}, {Barache}, {Barata}, {Barbier}, {Barblan},
  {Baroni}, {Barrado y Navascu{\'e}s}, {Barros}, {Barstow}, {Becciani},
  {Bellazzini}, {Bellei}, {Bello Garc{\'\i}a}, {Belokurov}, {Bendjoya},
  {Berihuete}, {Bianchi}, {Bienaym{\'e}}, {Billebaud}, {Blagorodnova},
  {Blanco-Cuaresma}, {Boch}, {Bombrun}, {Borrachero}, {Bouquillon}, {Bourda},
  {Bouy}, {Bragaglia}, {Breddels}, {Brouillet}, {Br{\"u}semeister},
  {Bucciarelli}, {Budnik}, {Burgess}, {Burgon}, {Burlacu}, {Busonero}, {Buzzi},
  {Caffau}, {Cambras}, {Campbell}, {Cancelliere}, {Cantat-Gaudin}, {Carlucci},
  {Carrasco}, {Castellani}, {Charlot}, {Charnas}, {Charvet}, {Chassat},
  {Chiavassa}, {Clotet}, {Cocozza}, {Collins}, {Collins}, {Costigan}, {Crifo},
  {Cross}, {Crosta}, {Crowley}, {Dafonte}, {Damerdji}, {Dapergolas}, {David},
  {David}, {De Cat}, {de Felice}, {de Laverny}, {De Luise}, {De March}, {de
  Martino}, {de Souza}, {Debosscher}, {del Pozo}, {Delbo}, {Delgado},
  {Delgado}, {di Marco}, {Di Matteo}, {Diakite}, {Distefano}, {Dolding}, {Dos
  Anjos}, {Drazinos}, {Dur{\'a}n}, {Dzigan}, {Ecale}, {Edvardsson}, {Enke},
  {Erdmann}, {Escolar}, {Espina}, {Evans}, {Eynard Bontemps}, {Fabre},
  {Fabrizio}, {Faigler}, {Falc{\~a}o}, {Farr{\`a}s Casas}, {Faye}, {Federici},
  {Fedorets}, {Fern{\'a}ndez-Hern{\'a}ndez}, {Fernique}, {Fienga}, {Figueras},
  {Filippi}, {Findeisen}, {Fonti}, {Fouesneau}, {Fraile}, {Fraser}, {Fuchs},
  {Furnell}, {Gai}, {Galleti}, {Galluccio}, {Garabato}, {Garc{\'\i}a-Sedano},
  {Gar{\'e}}, {Garofalo}, {Garralda}, {Gavras}, {Gerssen}, {Geyer}, {Gilmore},
  {Girona}, {Giuffrida}, {Gomes}, {Gonz{\'a}lez-Marcos},
  {Gonz{\'a}lez-N{\'u}{\~n}ez}, {Gonz{\'a}lez-Vidal}, {Granvik}, {Guerrier},
  {Guillout}, {Guiraud}, {G{\'u}rpide}, {Guti{\'e}rrez-S{\'a}nchez}, {Guy},
  {Haigron}, {Hatzidimitriou}, {Haywood}, {Heiter}, {Helmi}, {Hobbs},
  {Hofmann}, {Holl}, {Holland}, {Hunt}, {Hypki}, {Icardi}, {Irwin}, {Jevardat
  de Fombelle}, {Jofr{\'e}}, {Jonker}, {Jorissen}, {Julbe}, {Karampelas},
  {Kochoska}, {Kohley}, {Kolenberg}, {Kontizas}, {Koposov}, {Kordopatis},
  {Koubsky}, {Kowalczyk}, {Krone-Martins}, {Kudryashova}, {Kull}, {Bachchan},
  {Lacoste-Seris}, {Lanza}, {Lavigne}, {Le Poncin-Lafitte}, {Lebreton},
  {Lebzelter}, {Leccia}, {Leclerc}, {Lecoeur-Taibi}, {Lemaitre}, {Lenhardt},
  {Leroux}, {Liao}, {Licata}, {Lindstr{\o}m}, {Lister}, {Livanou}, {Lobel},
  {L{\"o}ffler}, {L{\'o}pez}, {Lopez-Lozano}, {Lorenz}, {Loureiro},
  {MacDonald}, {Magalh{\~a}es Fernandes}, {Managau}, {Mann}, {Mantelet},
  {Marchal}, {Marchant}, {Marconi}, {Marie}, {Marinoni}, {Marrese},
  {Marschalk{\'o}}, {Marshall}, {Mart{\'\i}n-Fleitas}, {Martino}, {Mary},
  {Matijevi{\v{c}}}, {Mazeh}, {McMillan}, {Messina}, {Mestre}, {Michalik},
  {Millar}, {Miranda}, {Molina}, {Molinaro}, {Molinaro}, {Moln{\'a}r},
  {Moniez}, {Montegriffo}, {Monteiro}, {Mor}, {Mora}, {Morbidelli}, {Morel},
  {Morgenthaler}, {Morley}, {Morris}, {Mulone}, {Muraveva}, {Musella},
  {Narbonne}, {Nelemans}, {Nicastro}, {Noval}, {Ord{\'e}novic},
  {Ordieres-Mer{\'e}}, {Osborne}, {Pagani}, {Pagano}, {Pailler}, {Palacin},
  {Palaversa}, {Parsons}, {Paulsen}, {Pecoraro}, {Pedrosa}, {Pentik{\"a}inen},
  {Pereira}, {Pichon}, {Piersimoni}, {Pineau}, {Plachy}, {Plum}, {Poujoulet},
  {Pr{\v{s}}a}, {Pulone}, {Ragaini}, {Rago}, {Rambaux}, {Ramos-Lerate},
  {Ranalli}, {Rauw}, {Read}, {Regibo}, {Renk}, {Reyl{\'e}}, {Ribeiro},
  {Rimoldini}, {Ripepi}, {Riva}, {Rixon}, {Roelens}, {Romero-G{\'o}mez},
  {Rowell}, {Royer}, {Rudolph}, {Ruiz-Dern}, {Sadowski}, {Sagrist{\`a}
  Sell{\'e}s}, {Sahlmann}, {Salgado}, {Salguero}, {Sarasso}, {Savietto},
  {Schnorhk}, {Schultheis}, {Sciacca}, {Segol}, {Segovia}, {Segransan},
  {Serpell}, {Shih}, {Smareglia}, {Smart}, {Smith}, {Solano}, {Solitro},
  {Sordo}, {Soria Nieto}, {Souchay}, {Spagna}, {Spoto}, {Stampa}, {Steele},
  {Steidelm{\"u}ller}, {Stephenson}, {Stoev}, {Suess}, {S{\"u}veges}, {Surdej},
  {Szabados}, {Szegedi-Elek}, {Tapiador}, {Taris}, {Tauran}, {Taylor},
  {Teixeira}, {Terrett}, {Tingley}, {Trager}, {Turon}, {Ulla}, {Utrilla},
  {Valentini}, {van Elteren}, {Van Hemelryck}, {van Leeuwen}, {Varadi},
  {Vecchiato}, {Veljanoski}, {Via}, {Vicente}, {Vogt}, {Voss}, {Votruba},
  {Voutsinas}, {Walmsley}, {Weiler}, {Weingrill}, {Werner}, {Wevers},
  {Whitehead}, {Wyrzykowski}, {Yoldas}, {{\v{Z}}erjal}, {Zucker}, {Zurbach},
  {Zwitter}, {Alecu}, {Allen}, {Allende Prieto}, {Amorim},
  {Anglada-Escud{\'e}}, {Arsenijevic}, {Azaz}, {Balm}, {Beck}, {Bernstein},
  {Bigot}, {Bijaoui}, {Blasco}, {Bonfigli}, {Bono}, {Boudreault}, {Bressan},
  {Brown}, {Brunet}, {Bunclark}, {Buonanno}, {Butkevich}, {Carret}, {Carrion},
  {Chemin}, {Ch{\'e}reau}, {Corcione}, {Darmigny}, {de Boer}, {de Teodoro}, {de
  Zeeuw}, {Delle Luche}, {Domingues}, {Dubath}, {Fodor}, {Fr{\'e}zouls},
  {Fries}, {Fustes}, {Fyfe}, {Gallardo}, {Gallegos}, {Gardiol}, {Gebran},
  {Gomboc}, {G{\'o}mez}, {Grux}, {Gueguen}, {Heyrovsky}, {Hoar}, {Iannicola},
  {Isasi Parache}, {Janotto}, {Joliet}, {Jonckheere}, {Keil}, {Kim},
  {Klagyivik}, {Klar}, {Knude}, {Kochukhov}, {Kolka}, {Kos}, {Kutka}, {Lainey},
  {LeBouquin}, {Liu}, {Loreggia}, {Makarov}, {Marseille}, {Martayan},
  {Martinez-Rubi}, {Massart}, {Meynadier}, {Mignot}, {Munari}, {Nguyen},
  {Nordlander}, {Ocvirk}, {O'Flaherty}, {Olias Sanz}, {Ortiz}, {Osorio},
  {Oszkiewicz}, {Ouzounis}, {Palmer}, {Park}, {Pasquato}, {Peltzer}, {Peralta},
  {P{\'e}turaud}, {Pieniluoma}, {Pigozzi}, {Poels}, {Prat}, {Prod'homme},
  {Raison}, {Rebordao}, {Risquez}, {Rocca-Volmerange}, {Rosen}, {Ruiz-Fuertes},
  {Russo}, {Sembay}, {Serraller Vizcaino}, {Short}, {Siebert}, {Silva},
  {Sinachopoulos}, {Slezak}, {Soffel}, {Sosnowska}, {Strai{\v{z}}ys}, {ter
  Linden}, {Terrell}, {Theil}, {Tiede}, {Troisi}, {Tsalmantza}, {Tur},
  {Vaccari}, {Vachier}, {Valles}, {Van Hamme}, {Veltz}, {Virtanen}, {Wallut},
  {Wichmann}, {Wilkinson}, {Ziaeepour}, \& {Zschocke}}]{2016A&A...595A...1G}
{Gaia Collaboration}, {Prusti}, T., {de Bruijne}, J.~H.~J., {et~al.} 2016,
  \aap, 595, A1

\bibitem[{George \& Huerta(2018)}]{George_2018}
George, D. \& Huerta, E. 2018, Physics Letters B, 778, 64–70

\bibitem[{{Hendriks} \& {Aerts}(2019)}]{2019PASP..131j8001H}
{Hendriks}, L. \& {Aerts}, C. 2019, \pasp, 131, 108001

\bibitem[{Hoeser \& Kuenzer(2020)}]{Hoeser_2020}
Hoeser, T. \& Kuenzer, C. 2020, Remote Sensing, 12, 1667

\bibitem[{{Hojjatpanah} {et~al.}(2019){Hojjatpanah}, {Figueira}, {Santos},
  {Adibekyan}, {Sousa}, {Delgado-Mena}, {Alibert}, {Cristiani}, {Gonz{\'a}lez
  Hern{\'a}ndez}, {Lanza}, {Di Marcantonio}, {Martins}, {Micela}, {Molaro},
  {Neves}, {Oshagh}, {Pepe}, {Poretti}, {Rojas-Ayala}, {Rebolo}, {Su{\'a}rez
  Mascare{\~n}o}, \& {Zapatero Osorio}}]{2019A&A...629A..80H}
{Hojjatpanah}, S., {Figueira}, P., {Santos}, N.~C., {et~al.} 2019, \aap, 629,
  A80

\bibitem[{{Howarth} {et~al.}(1997){Howarth}, {Siebert}, {Hussain}, \&
  {Prinja}}]{1997MNRAS.284..265H}
{Howarth}, I.~D., {Siebert}, K.~W., {Hussain}, G. A.~J., \& {Prinja}, R.~K.
  1997, \mnras, 284, 265

\bibitem[{Ivezić {et~al.}(2019)Ivezić, Kahn, Tyson, Abel, Acosta, Allsman,
  Alonso, AlSayyad, Anderson, Andrew, \& et~al.}]{Ivezi_2019}
Ivezić, v., Kahn, S.~M., Tyson, J.~A., {et~al.} 2019, The Astrophysical
  Journal, 873, 111

\bibitem[{Kingma \& Ba(2017)}]{kingma2017adam}
Kingma, D.~P. \& Ba, J. 2017, Adam: A Method for Stochastic Optimization

\bibitem[{Kirillov {et~al.}(2019)Kirillov, Girshick, He, \&
  Dollár}]{kirillov2019panoptic}
Kirillov, A., Girshick, R., He, K., \& Dollár, P. 2019, Panoptic Feature
  Pyramid Networks

\bibitem[{{Kuka{\v{c}}ka} {et~al.}(2017){Kuka{\v{c}}ka}, {Golkov}, \&
  {Cremers}}]{2017arXiv171010686K}
{Kuka{\v{c}}ka}, J., {Golkov}, V., \& {Cremers}, D. 2017, arXiv e-prints,
  arXiv:1710.10686

\bibitem[{Kurucz(1970)}]{kurucz1970atlas}
Kurucz, R.~L. 1970, SAO Special report, 309

\bibitem[{{Lanz} \& {Hubeny}(2003)}]{2003ApJS..146..417L}
{Lanz}, T. \& {Hubeny}, I. 2003, \apjs, 146, 417

\bibitem[{{Lanz} \& {Hubeny}(2007)}]{2007ApJS..169...83L}
{Lanz}, T. \& {Hubeny}, I. 2007, \apjs, 169, 83

\bibitem[{LeCun {et~al.}(1989)LeCun, Boser, Denker, Henderson, Howard, Hubbard,
  \& Jackel}]{lecun_1989}
LeCun, Y., Boser, B., Denker, J.~S., {et~al.} 1989, Neural Computation, 1, 541

\bibitem[{Lin {et~al.}(2017)Lin, Dollár, Girshick, He, Hariharan, \&
  Belongie}]{lin2017feature}
Lin, T.-Y., Dollár, P., Girshick, R., {et~al.} 2017, Feature Pyramid Networks
  for Object Detection

\bibitem[{Long {et~al.}(2015)Long, Shelhamer, \& Darrell}]{long2015fully}
Long, J., Shelhamer, E., \& Darrell, T. 2015, Fully Convolutional Networks for
  Semantic Segmentation

\bibitem[{{Mahabal} {et~al.}(2019){Mahabal}, {Rebbapragada}, {Walters},
  {Masci}, {Blagorodnova}, {van Roestel}, {Ye}, {Biswas}, {Burdge}, {Chang},
  {Duev}, {Golkhou}, {Miller}, {Nordin}, {Ward}, {Adams}, {Bellm}, {Branton},
  {Bue}, {Cannella}, {Connolly}, {Dekany}, {Feindt}, {Hung}, {Fortson},
  {Frederick}, {Fremling}, {Gezari}, {Graham}, {Groom}, {Kasliwal}, {Kulkarni},
  {Kupfer}, {Lin}, {Lintott}, {Lunnan}, {Parejko}, {Prince}, {Riddle},
  {Rusholme}, {Saunders}, {Sedaghat}, {Shupe}, {Singer}, {Soumagnac}, {Szkody},
  {Tachibana}, {Tirumala}, {van Velzen}, \& {Wright}}]{2019PASP..131c8002M}
{Mahabal}, A., {Rebbapragada}, U., {Walters}, R., {et~al.} 2019, \pasp, 131,
  038002

\bibitem[{{Majewski} {et~al.}(2017){Majewski}, {Schiavon}, {Frinchaboy},
  {Allende Prieto}, {Barkhouser}, {Bizyaev}, {Blank}, {Brunner}, {Burton},
  {Carrera}, {Chojnowski}, {Cunha}, {Epstein}, {Fitzgerald}, {Garc{\'\i}a
  P{\'e}rez}, {Hearty}, {Henderson}, {Holtzman}, {Johnson}, {Lam}, {Lawler},
  {Maseman}, {M{\'e}sz{\'a}ros}, {Nelson}, {Nguyen}, {Nidever}, {Pinsonneault},
  {Shetrone}, {Smee}, {Smith}, {Stolberg}, {Skrutskie}, {Walker}, {Wilson},
  {Zasowski}, {Anders}, {Basu}, {Beland}, {Blanton}, {Bovy}, {Brownstein},
  {Carlberg}, {Chaplin}, {Chiappini}, {Eisenstein}, {Elsworth}, {Feuillet},
  {Fleming}, {Galbraith-Frew}, {Garc{\'\i}a}, {Garc{\'\i}a-Hern{\'a}ndez},
  {Gillespie}, {Girardi}, {Gunn}, {Hasselquist}, {Hayden}, {Hekker}, {Ivans},
  {Kinemuchi}, {Klaene}, {Mahadevan}, {Mathur}, {Mosser}, {Muna}, {Munn},
  {Nichol}, {O'Connell}, {Parejko}, {Robin}, {Rocha-Pinto}, {Schultheis},
  {Serenelli}, {Shane}, {Silva Aguirre}, {Sobeck}, {Thompson}, {Troup},
  {Weinberg}, \& {Zamora}}]{2017AJ....154...94M}
{Majewski}, S.~R., {Schiavon}, R.~P., {Frinchaboy}, P.~M., {et~al.} 2017, \aj,
  154, 94

\bibitem[{{Negueruela} {et~al.}(2004){Negueruela}, {Steele}, \&
  {Bernabeu}}]{2004AN....325..749N}
{Negueruela}, I., {Steele}, I.~A., \& {Bernabeu}, G. 2004, Astronomische
  Nachrichten, 325, 749

\bibitem[{Nesterov(1983)}]{nesterov1983method}
Nesterov, Y.~E. 1983, in Dokl. akad. nauk Sssr, Vol. 269, 543--547

\bibitem[{Newell {et~al.}(2016)Newell, Yang, \& Deng}]{newell2016stacked}
Newell, A., Yang, K., \& Deng, J. 2016, Stacked Hourglass Networks for Human
  Pose Estimation

\bibitem[{{Nissen} {et~al.}(2020){Nissen}, {Christensen-Dalsgaard},
  {Mosumgaard}, {Silva Aguirre}, {Spitoni}, \& {Verma}}]{Nissen_2020}
{Nissen}, P.~E., {Christensen-Dalsgaard}, J., {Mosumgaard}, J.~R., {et~al.}
  2020, \aap, 640, A81

\bibitem[{Noh {et~al.}(2015)Noh, Hong, \& Han}]{noh2015learning}
Noh, H., Hong, S., \& Han, B. 2015, Learning Deconvolution Network for Semantic
  Segmentation

\bibitem[{{Radosavovic} {et~al.}(2020){Radosavovic}, {Prateek Kosaraju},
  {Girshick}, {He}, \& {Doll{\'a}r}}]{2020arXiv200313678R}
{Radosavovic}, I., {Prateek Kosaraju}, R., {Girshick}, R., {He}, K., \&
  {Doll{\'a}r}, P. 2020, arXiv e-prints, arXiv:2003.13678

\bibitem[{Ronneberger {et~al.}(2015)Ronneberger, Fischer, \&
  Brox}]{ronneberger2015unet}
Ronneberger, O., Fischer, P., \& Brox, T. 2015, U-Net: Convolutional Networks
  for Biomedical Image Segmentation

\bibitem[{{Royer}(2009)}]{2009LNP...765..207R}
{Royer}, F. 2009, {On the Rotation of A-Type Stars}, Vol. 765, 207--230

\bibitem[{{Savitzky} \& {Golay}(1964)}]{1964AnaCh..36.1627S}
{Savitzky}, A. \& {Golay}, M.~J.~E. 1964, Analytical Chemistry, 36, 1627

\bibitem[{Scherer {et~al.}(2010)Scherer, M{\"u}ller, \&
  Behnke}]{scherer2010evaluation}
Scherer, D., M{\"u}ller, A., \& Behnke, S. 2010, in International conference on
  artificial neural networks, Springer, 92--101

\bibitem[{{Schr{\"o}der} {et~al.}(2009){Schr{\"o}der}, {Reiners}, \&
  {Schmitt}}]{Schroder_2009}
{Schr{\"o}der}, C., {Reiners}, A., \& {Schmitt}, J.~H.~M.~M. 2009, \aap, 493,
  1099

\bibitem[{Simonyan \& Zisserman(2015)}]{simonyan2015deep}
Simonyan, K. \& Zisserman, A. 2015, Very Deep Convolutional Networks for
  Large-Scale Image Recognition

\bibitem[{{Swihart} {et~al.}(2017){Swihart}, {Garcia}, {Stassun}, {van Belle},
  {Mutterspaugh}, \& {Elias}}]{2017AJ....153...16S}
{Swihart}, S.~J., {Garcia}, E.~V., {Stassun}, K.~G., {et~al.} 2017, \aj, 153,
  16

\bibitem[{Virtanen {et~al.}(2020)Virtanen, Gommers, Oliphant, Haberland, Reddy,
  Cournapeau, Burovski, Peterson, Weckesser, Bright, {van der Walt}, Brett,
  Wilson, Millman, Mayorov, Nelson, Jones, Kern, Larson, Carey, Polat, Feng,
  Moore, {VanderPlas}, Laxalde, Perktold, Cimrman, Henriksen, Quintero, Harris,
  Archibald, Ribeiro, Pedregosa, {van Mulbregt}, \& {SciPy 1.0
  Contributors}}]{2020SciPy-NMeth}
Virtanen, P., Gommers, R., Oliphant, T.~E., {et~al.} 2020, Nature Methods, 17,
  261

\bibitem[{{{\v{S}}koda} {et~al.}(2020){{\v{S}}koda}, {Podsztavek}, \&
  {Tvrd{\'\i}k}}]{2020A&A...643A.122S}
{{\v{S}}koda}, P., {Podsztavek}, O., \& {Tvrd{\'\i}k}, P. 2020, \aap, 643, A122

\bibitem[{{Walborn}(1973)}]{1973AJ.....78.1067W}
{Walborn}, N.~R. 1973, \aj, 78, 1067

\bibitem[{{Xie} {et~al.}(2016){Xie}, {Girshick}, {Doll{\'a}r}, {Tu}, \&
  {He}}]{2016arXiv161105431X}
{Xie}, S., {Girshick}, R., {Doll{\'a}r}, P., {Tu}, Z., \& {He}, K. 2016, arXiv
  e-prints, arXiv:1611.05431

\bibitem[{Xu {et~al.}(2019)Xu, Cisewski-Kehe, Davis, Fischer, \&
  Brewer}]{Xu_2019}
Xu, X., Cisewski-Kehe, J., Davis, A.~B., Fischer, D.~A., \& Brewer, J.~M. 2019,
  The Astronomical Journal, 157, 243

\bibitem[{Zhao {et~al.}(2012)Zhao, Zhao, Chu, Jing, \& Deng}]{zhao2012lamost}
Zhao, G., Zhao, Y., Chu, Y., Jing, Y., \& Deng, L. 2012, LAMOST Spectral Survey

\bibitem[{Zhao {et~al.}(2017)Zhao, Shi, Qi, Wang, \& Jia}]{zhao2017pyramid}
Zhao, H., Shi, J., Qi, X., Wang, X., \& Jia, J. 2017, Pyramid Scene Parsing
  Network

\bibitem[{Zhou {et~al.}(2018)Zhou, Siddiquee, Tajbakhsh, \&
  Liang}]{zhou2018unet}
Zhou, Z., Siddiquee, M. M.~R., Tajbakhsh, N., \& Liang, J. 2018, UNet++: A
  Nested U-Net Architecture for Medical Image Segmentation

\bibitem[{{Zorec} \& {Royer}(2012)}]{2012A&A...537A.120Z}
{Zorec}, J. \& {Royer}, F. 2012, \aap, 537, A120

\end{thebibliography}
%


\begin{appendix} 

\section{Diagrams of neural networks architectures}
\label{appendix:diagrams_of_nn}

Block diagrams of neural network architectures included in exploratory tests.

\begin{figure}
\resizebox{\hsize}{!}
{\includegraphics[clip]{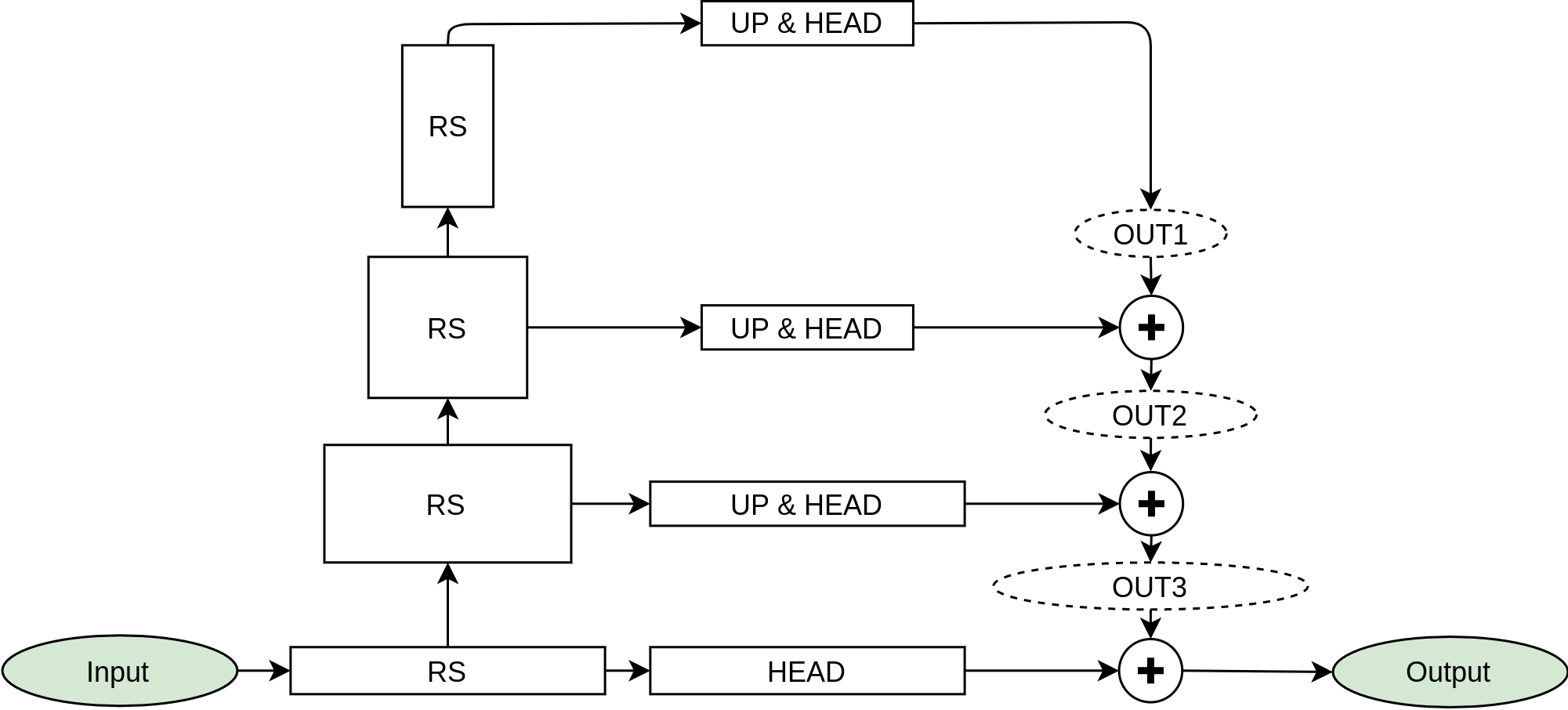}}
\caption{Fully Convolutional Network \citep[FCN,][]{long2015fully}}
\label{fig:all_nets-FCNet}
\end{figure}

\begin{figure}
\resizebox{\hsize}{!}
{\includegraphics[clip]{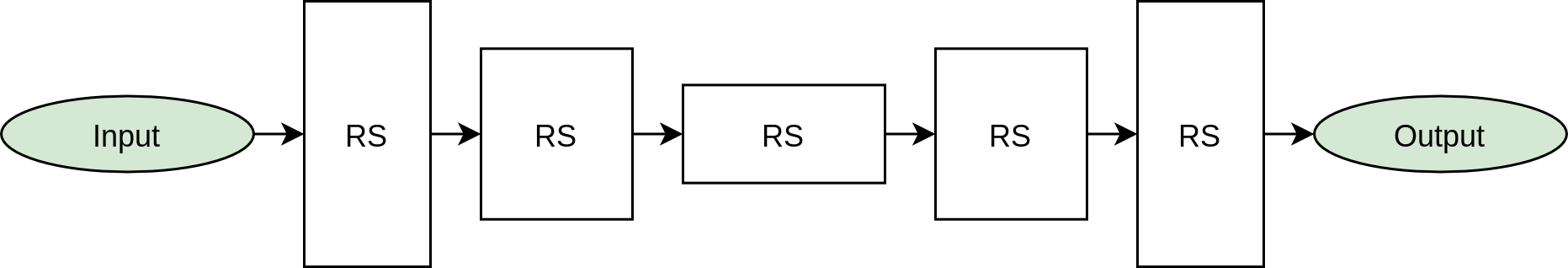}}
\caption{Deconvolution Network \citep[DeconvNet,][]{noh2015learning}}
\label{fig:all_nets-DeconvNet}
\end{figure}

\begin{figure}
\resizebox{\hsize}{!}
{\includegraphics[clip]{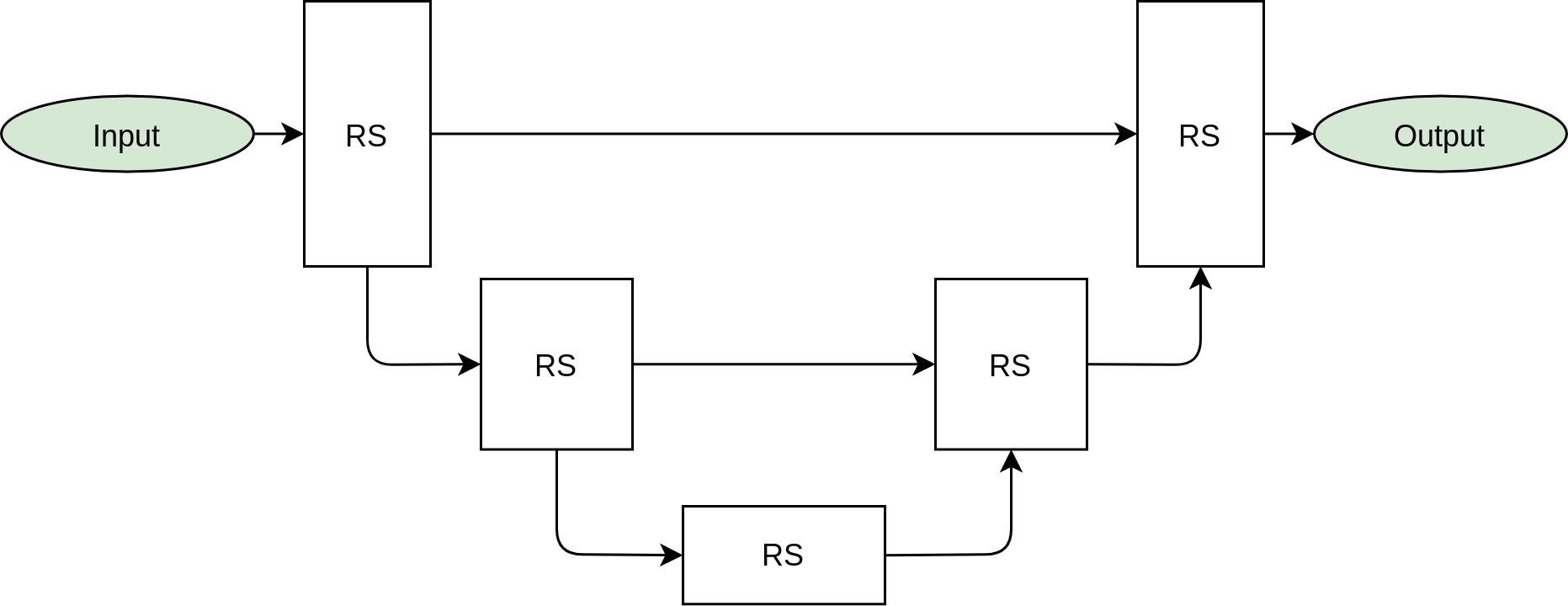}}
\caption{U-Net \citep{ronneberger2015unet}}
\label{fig:all_nets-U-Net}
\end{figure}

\begin{figure}
\resizebox{\hsize}{!}
{\includegraphics[clip]{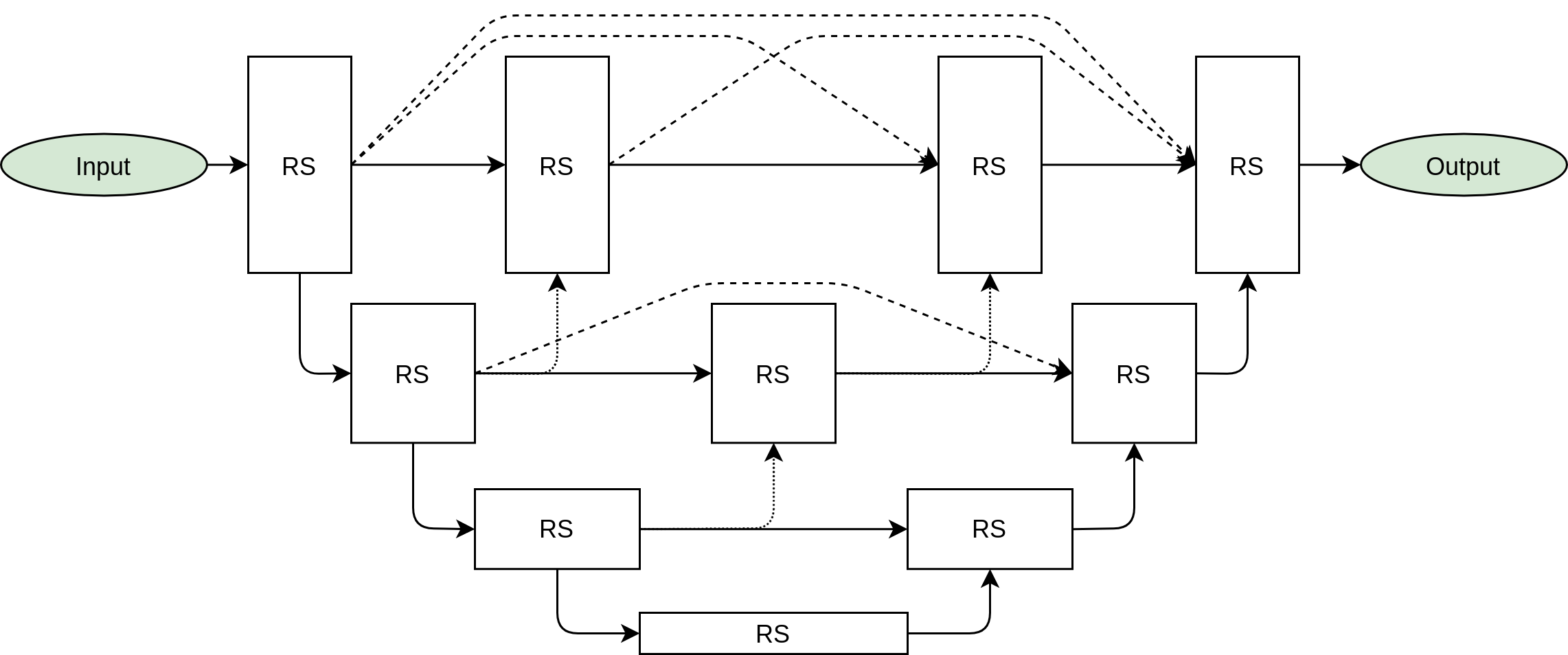}}
\caption{UNet++ \citep{zhou2018unet}}
\label{fig:all_nets-UNetpp}
\end{figure}

\begin{figure}
\resizebox{\hsize}{!}
{\includegraphics[clip]{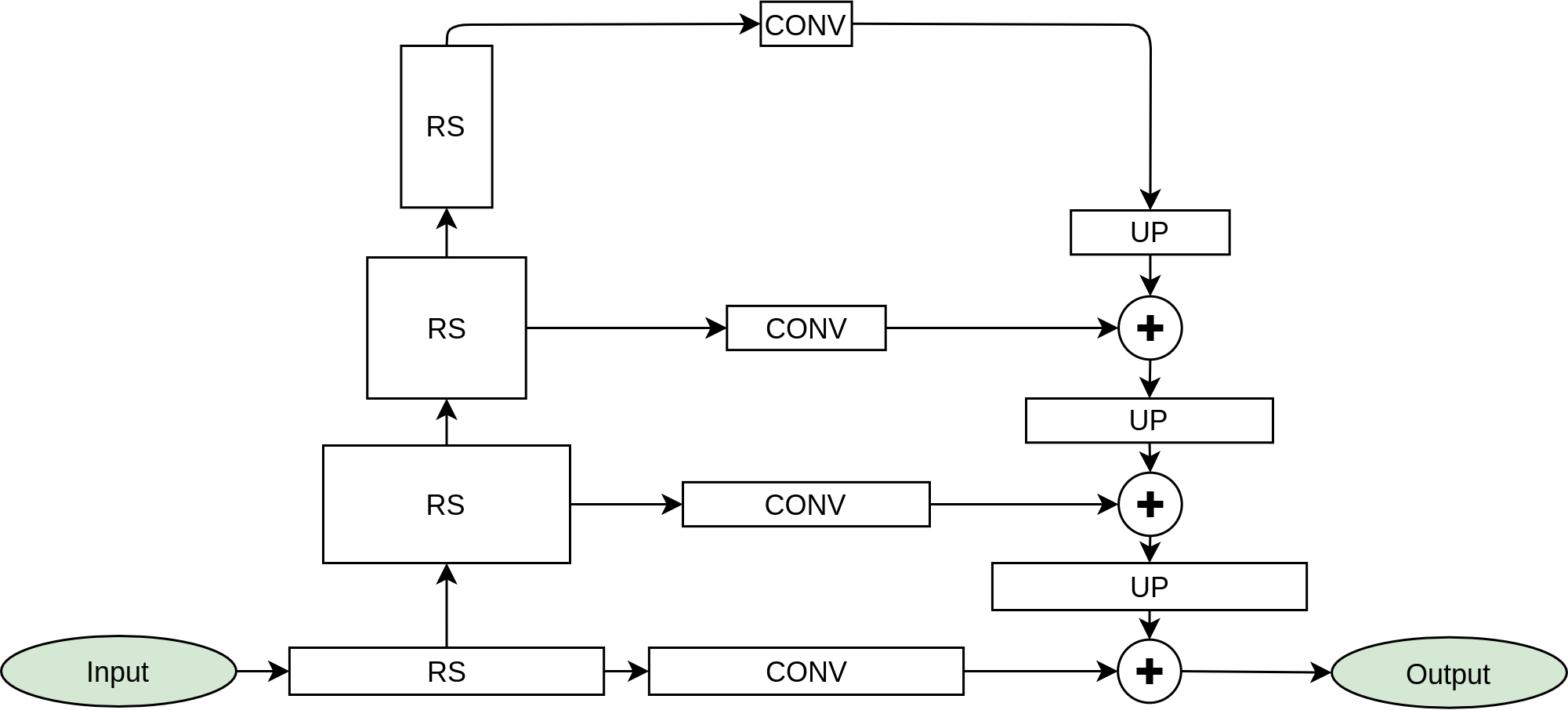}}
\caption{Feature Pyramid Network \citep[FPN,][]{lin2017feature, kirillov2019panoptic}}
\label{fig:all_nets-FPNet}
\end{figure}

\begin{figure}
\resizebox{\hsize}{!}
{\includegraphics[clip]{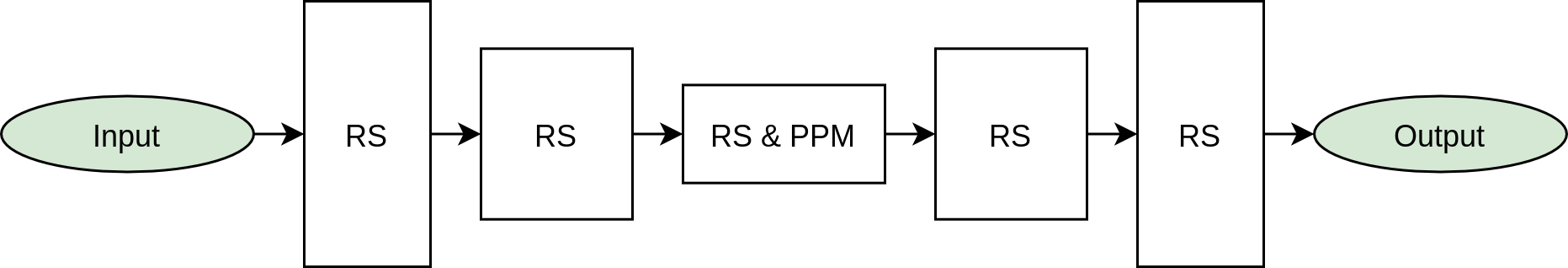}}
\caption{Pyramid Scene Parsing Network \citep[PSPNet,][]{zhao2017pyramid}}
\label{fig:all_nets-PSPNet}
\end{figure}

\begin{figure}
\resizebox{\hsize}{!}
{\includegraphics[clip]{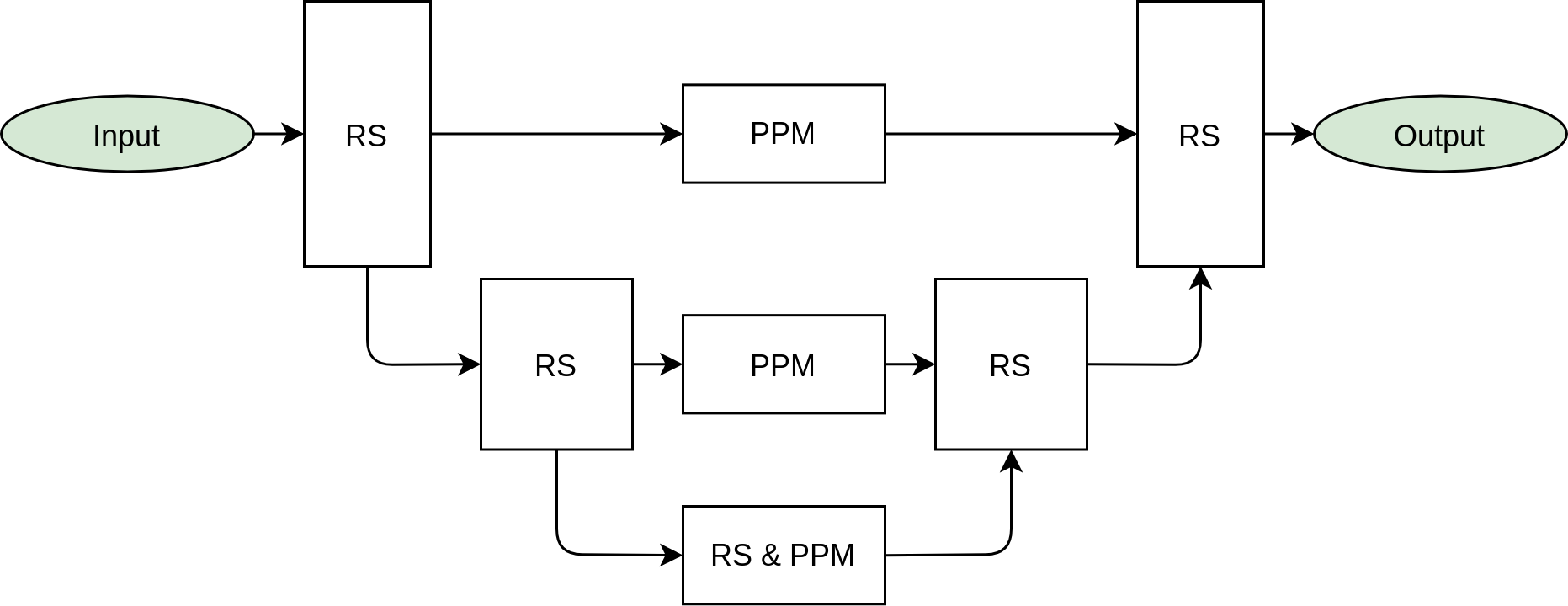}}
\caption{ U-Net with Pyramid Pooling Module (UPPNet, this work)}
\label{fig:all_nets-UPPNet}
\end{figure}

\section{Plots}
\label{appendix:graphs}
Detailed plots of the results.

\begin{figure*}
\resizebox{\hsize}{!}
{\includegraphics[clip]{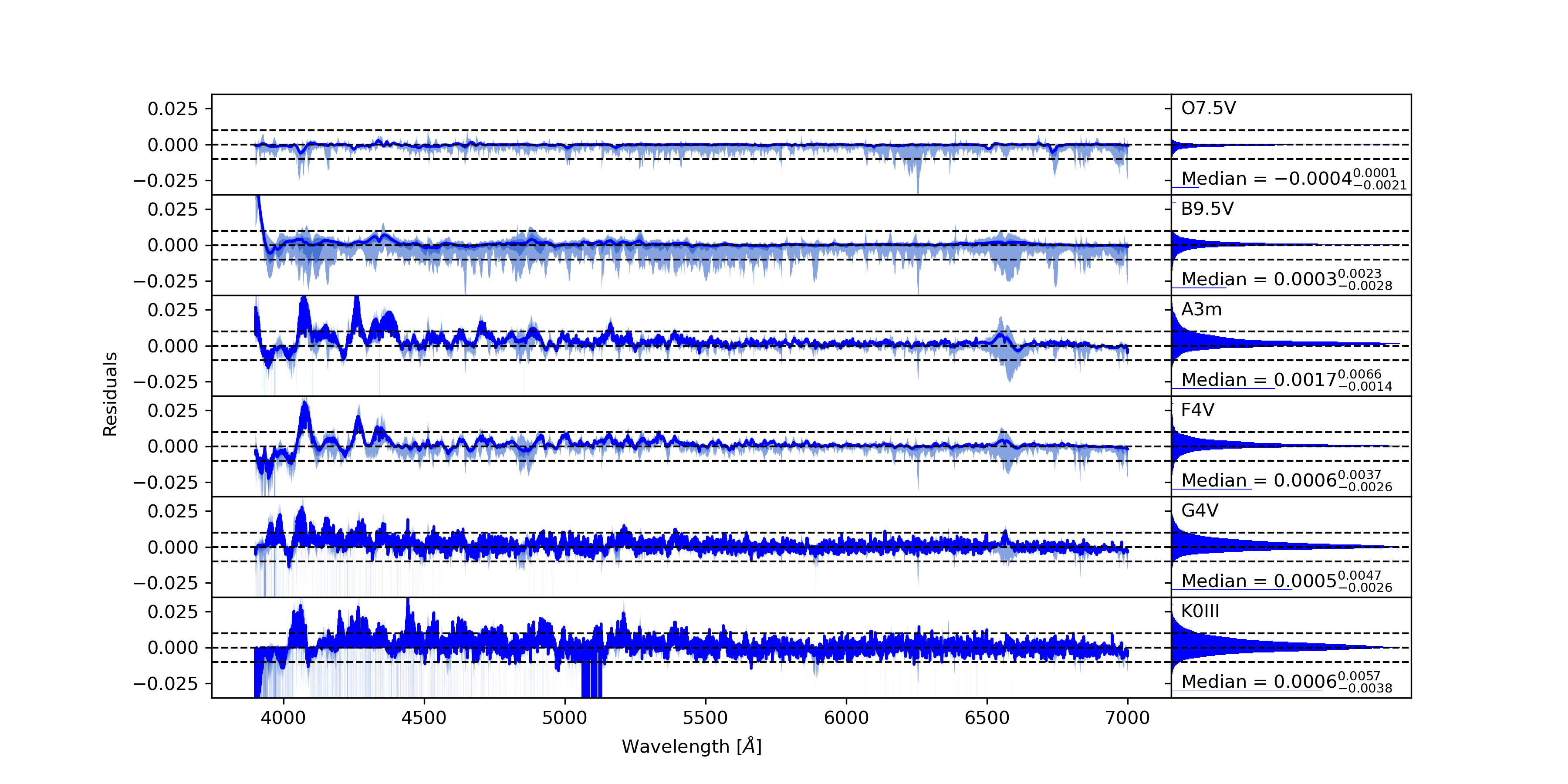}}
\caption{Results of normalisation of six synthetic spectra multiplied by six manually fitted \textit{pseudo-continua} using a neural network trained only with \textbf{synthetic data}. In each row, on the left, the differences between automatically normalised spectra and synthetic spectrum are shown, and on the right, the histogram of those differences with a related spectral type, the median with 15.87 percentile in the upper index, and 84.13 percentile in the lower index is displayed. The dashed lines on each panel correspond to the residuals equal -0.01, 0.0 and 0.01 respectively.}
\label{fig:synthetic_result_synth}
\end{figure*}

\begin{figure}
\resizebox{\hsize}{!}
{\includegraphics[clip]{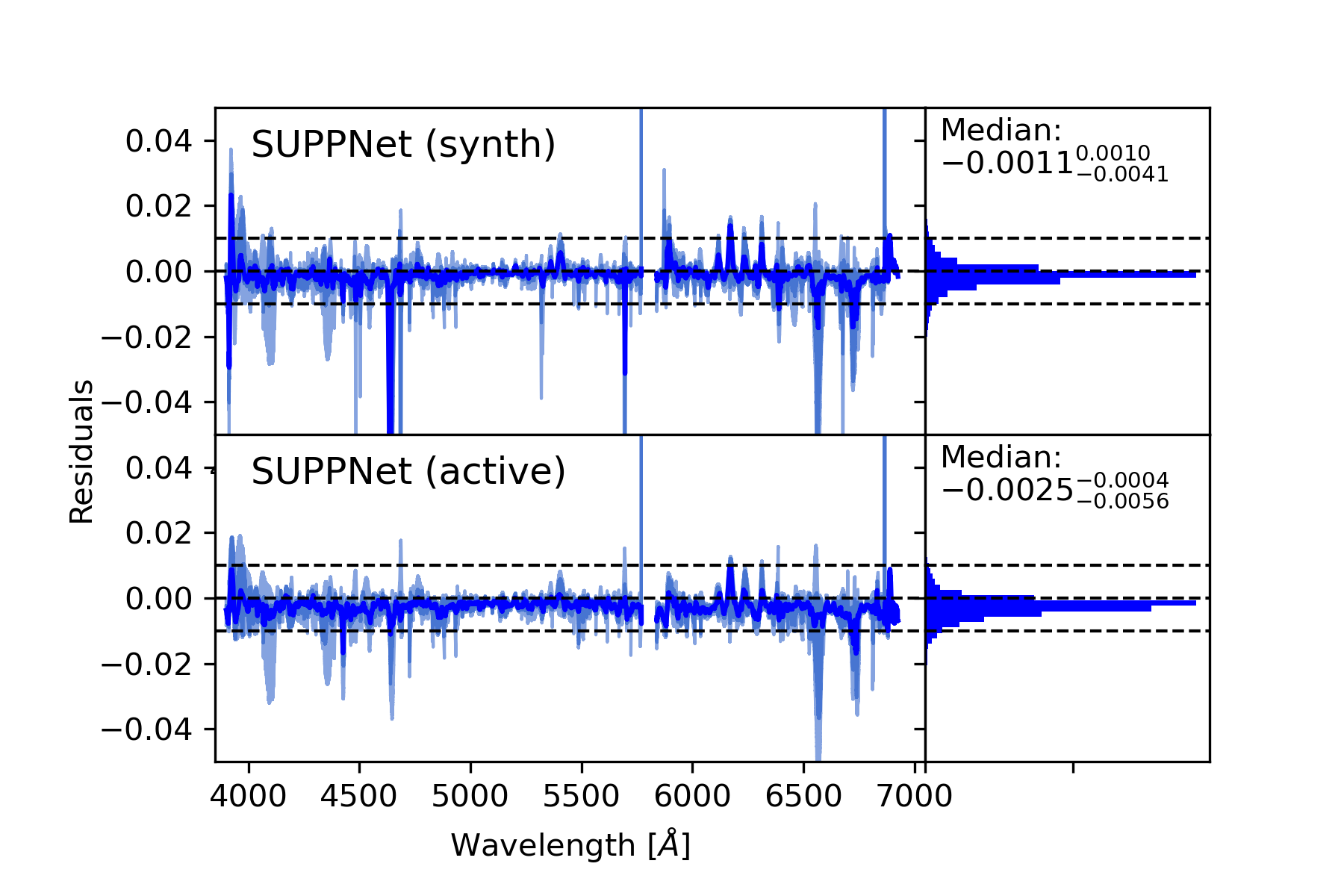}}
\caption{Residuals between the manually normalised spectrum and the result of the tested algorithm over \textbf{O type} stars from UVES POP field stars, that were manually normalised.}
\label{fig:UVES_POP_percentiles_hist_typeO}
\end{figure}

\begin{figure}
\resizebox{\hsize}{!}
{\includegraphics[clip]{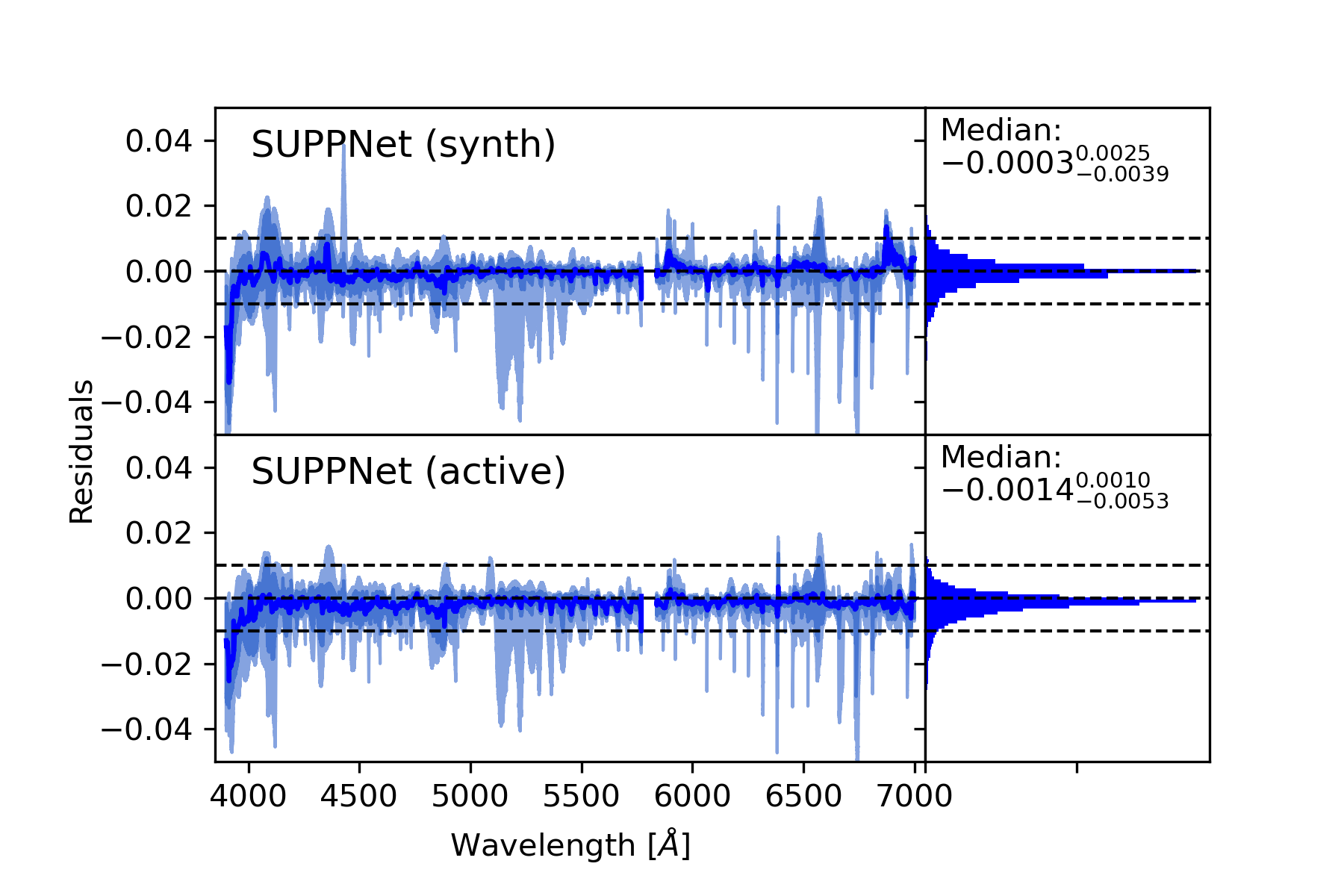}}
\caption{Residuals between the manually normalised spectrum and the result of the tested algorithm over \textbf{B type} stars from UVES POP field stars, that were manually normalised.}
\label{fig:UVES_POP_percentiles_hist_typeB}
\end{figure}

\begin{figure}
\resizebox{\hsize}{!}
{\includegraphics[clip]{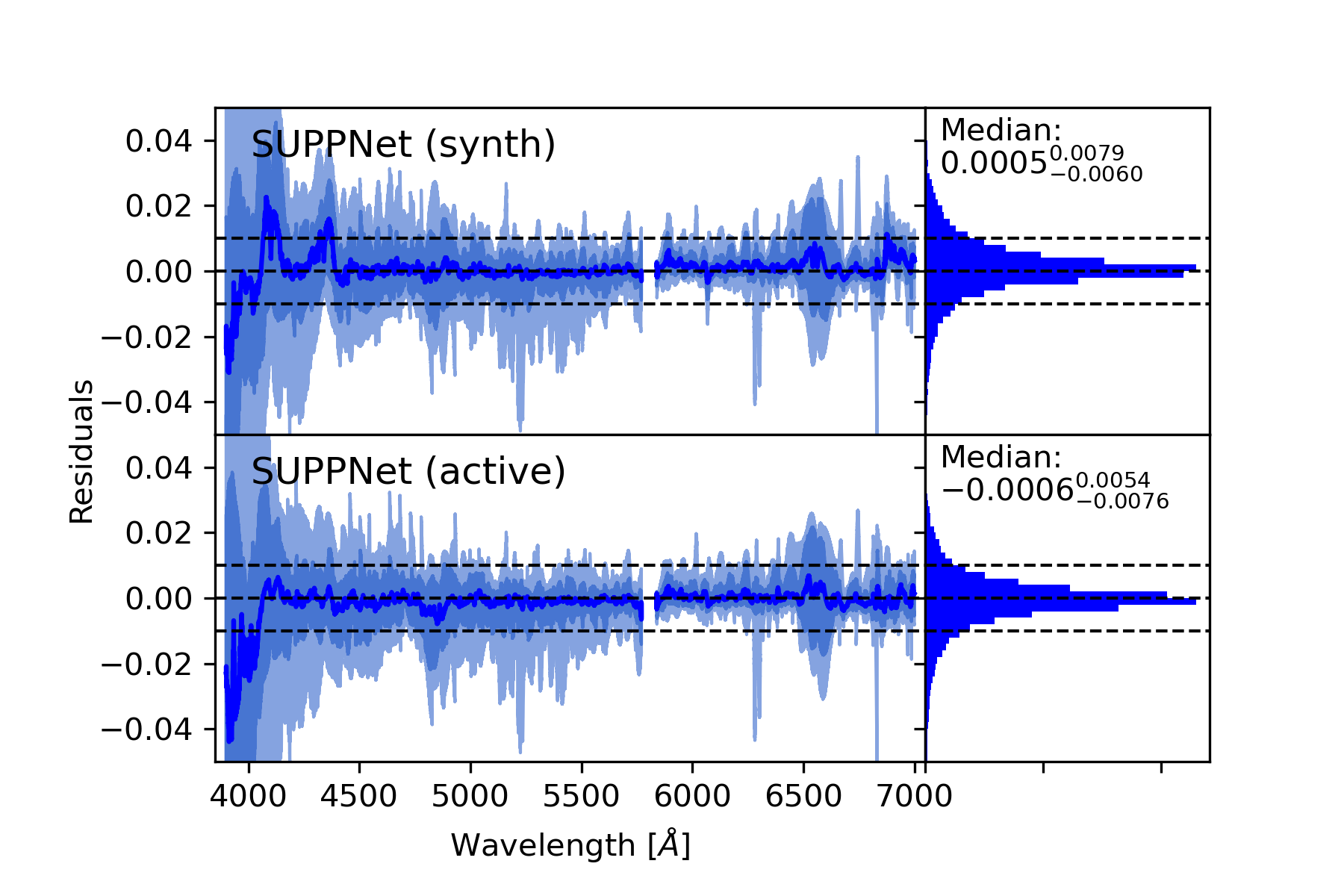}}
\caption{Residuals between the manually normalised spectrum and the result of the tested algorithm over \textbf{A type} stars from UVES POP field stars, that were manually normalised.}
\label{fig:UVES_POP_percentiles_hist_typeA}
\end{figure}

\begin{figure}
\resizebox{\hsize}{!}
{\includegraphics[clip]{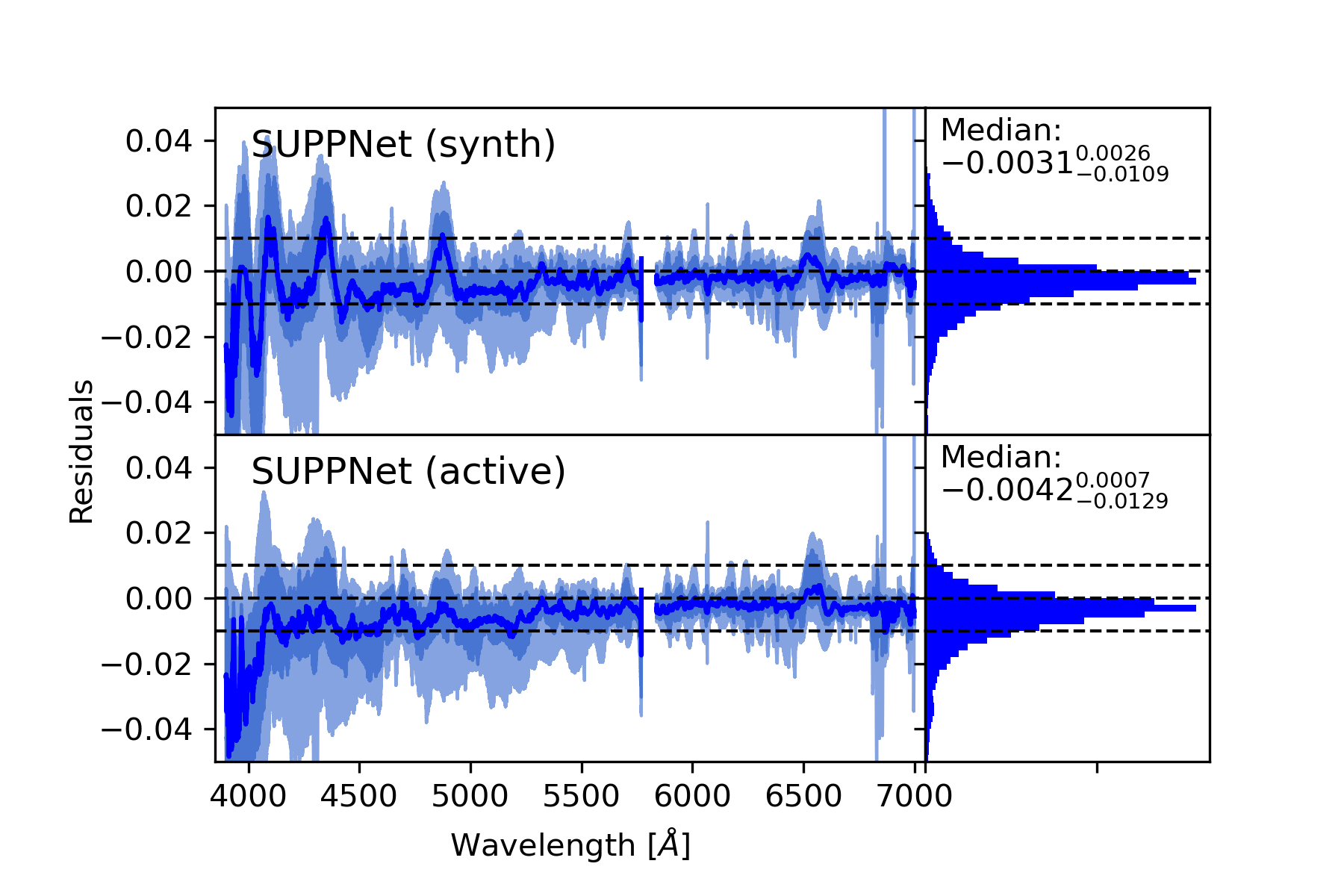}}
\caption{Residuals between the manually normalised spectrum and the result of the tested algorithm over \textbf{F type} stars from UVES POP field stars, that were manually normalised.}
\label{fig:UVES_POP_percentiles_hist_typeF}
\end{figure}

\begin{figure}
\resizebox{\hsize}{!}
{\includegraphics[clip]{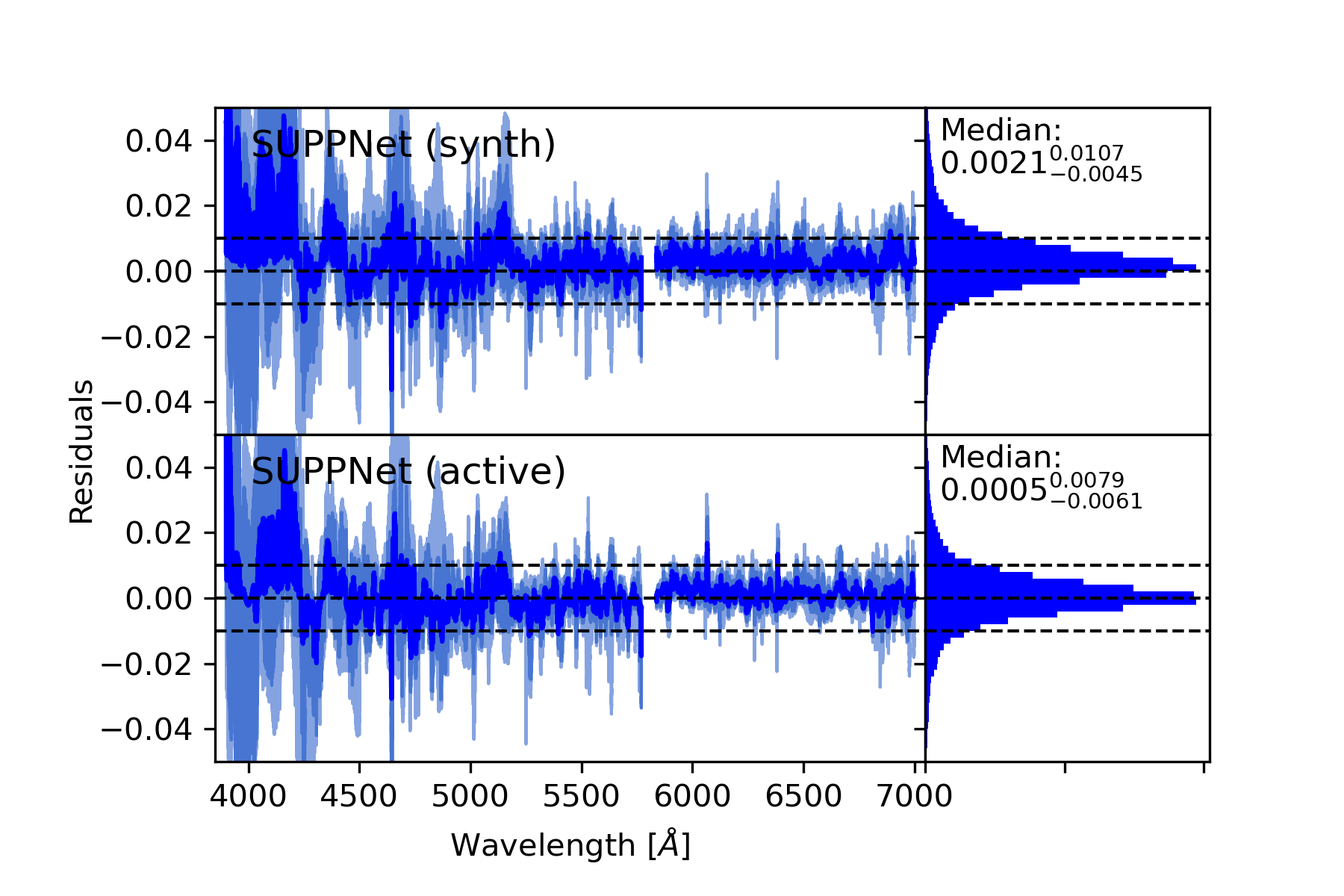}}
\caption{Residuals between the manually normalised spectrum and the result of the tested algorithm over \textbf{G type} stars from UVES POP field stars, that were manually normalised.}
\label{fig:UVES_POP_percentiles_hist_typeG}
\end{figure}

\begin{figure*}
\resizebox{\hsize}{!}
{\includegraphics[clip]{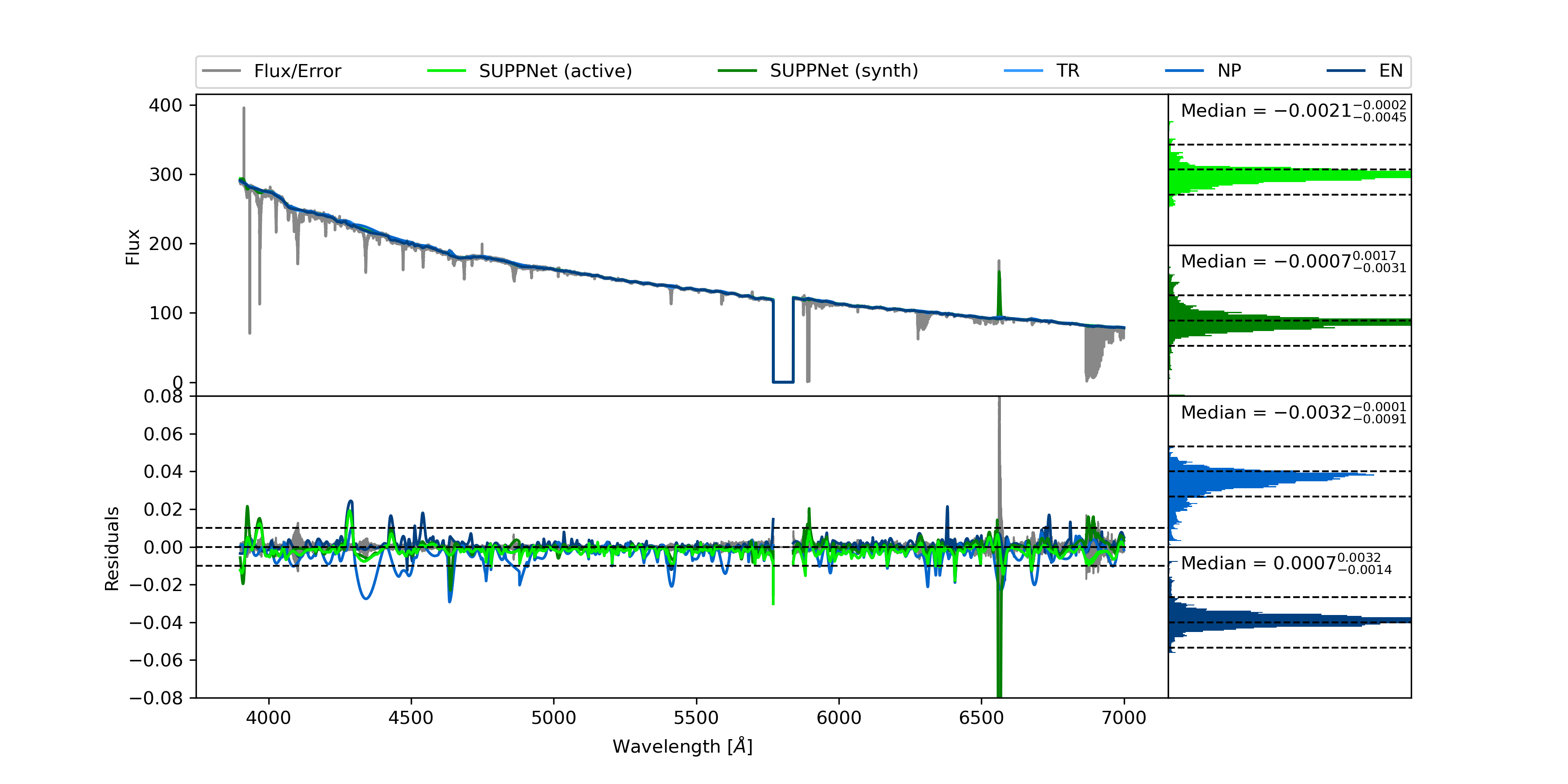}}
\caption{Comparison of normalisation quality on the example of \textbf{HD\,155806 (O7.5\,V)} star with two versions of the proposed method (SUPPNet active and synth) and manual normalisation done independently by three different people (TR, NP, and EN). The left upper panel shows original flux with all fitted \textit{pseudo-continua}. The left lower panel shows residuals of normalised fluxes relative to TR normalisation. The right panel presents histograms of all mentioned residuals with median, 15.87, and 84.13 percentiles.}
\label{fig:155806_manual}
\end{figure*}

\begin{figure*}
\resizebox{\hsize}{!}
{\includegraphics[clip]{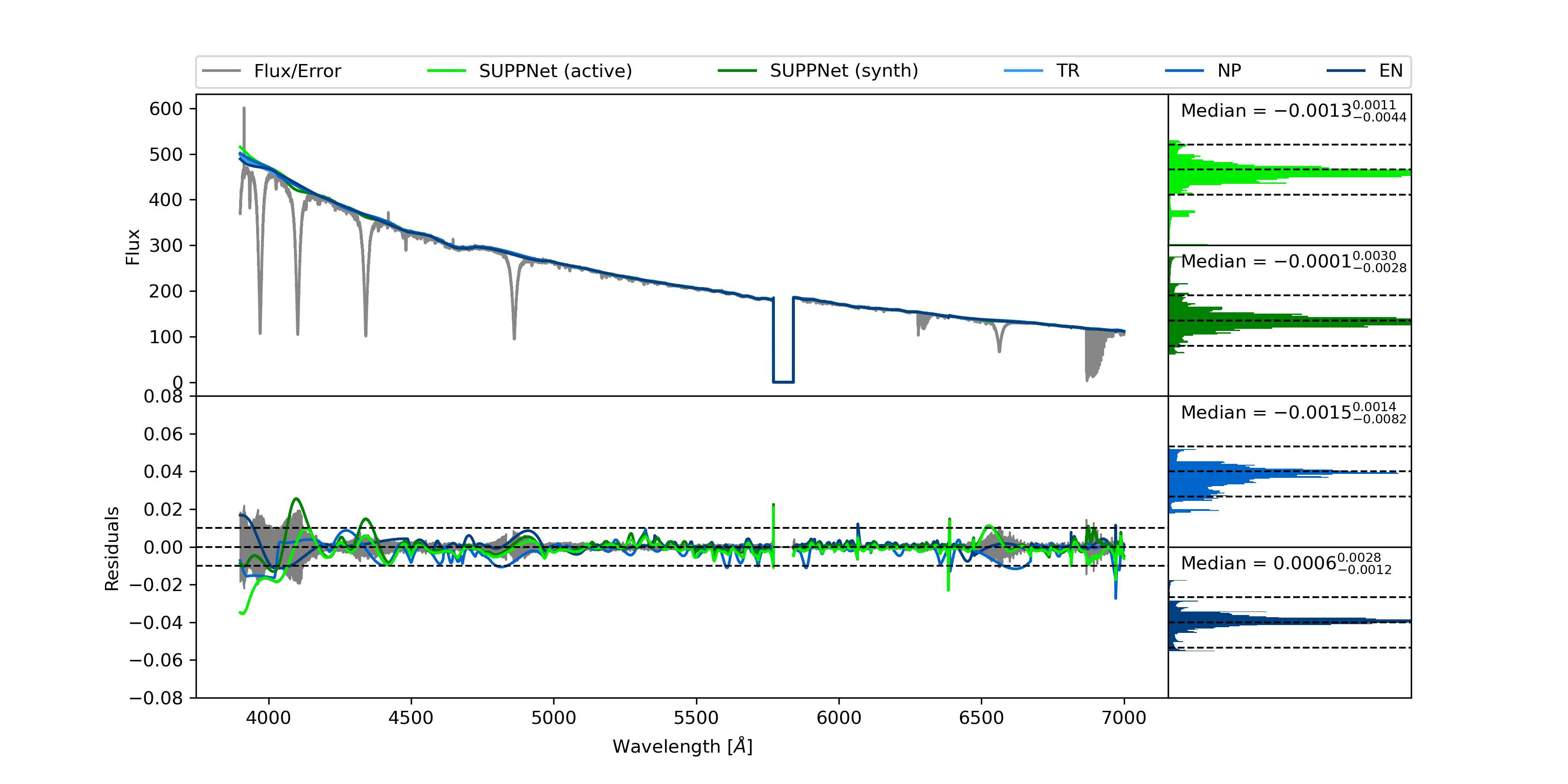}}
\caption{Comparison of normalisation quality on the example of \textbf{HD\,90882 (B9.5\,V)} star with two versions of the proposed method (SUPPNet active and synth) and manual normalisation done independently by three different people (TR, NP, and EN). The left upper panel shows original flux with all fitted \textit{pseudo-continua}. The left lower panel shows residuals of normalised fluxes relative to TR normalisation. The right panel presents histograms of all mentioned residuals with median, 15.87, and 84.13 percentiles.}
\label{fig:90882_manual}
\end{figure*}

\begin{figure*}
\resizebox{\hsize}{!}
{\includegraphics[clip]{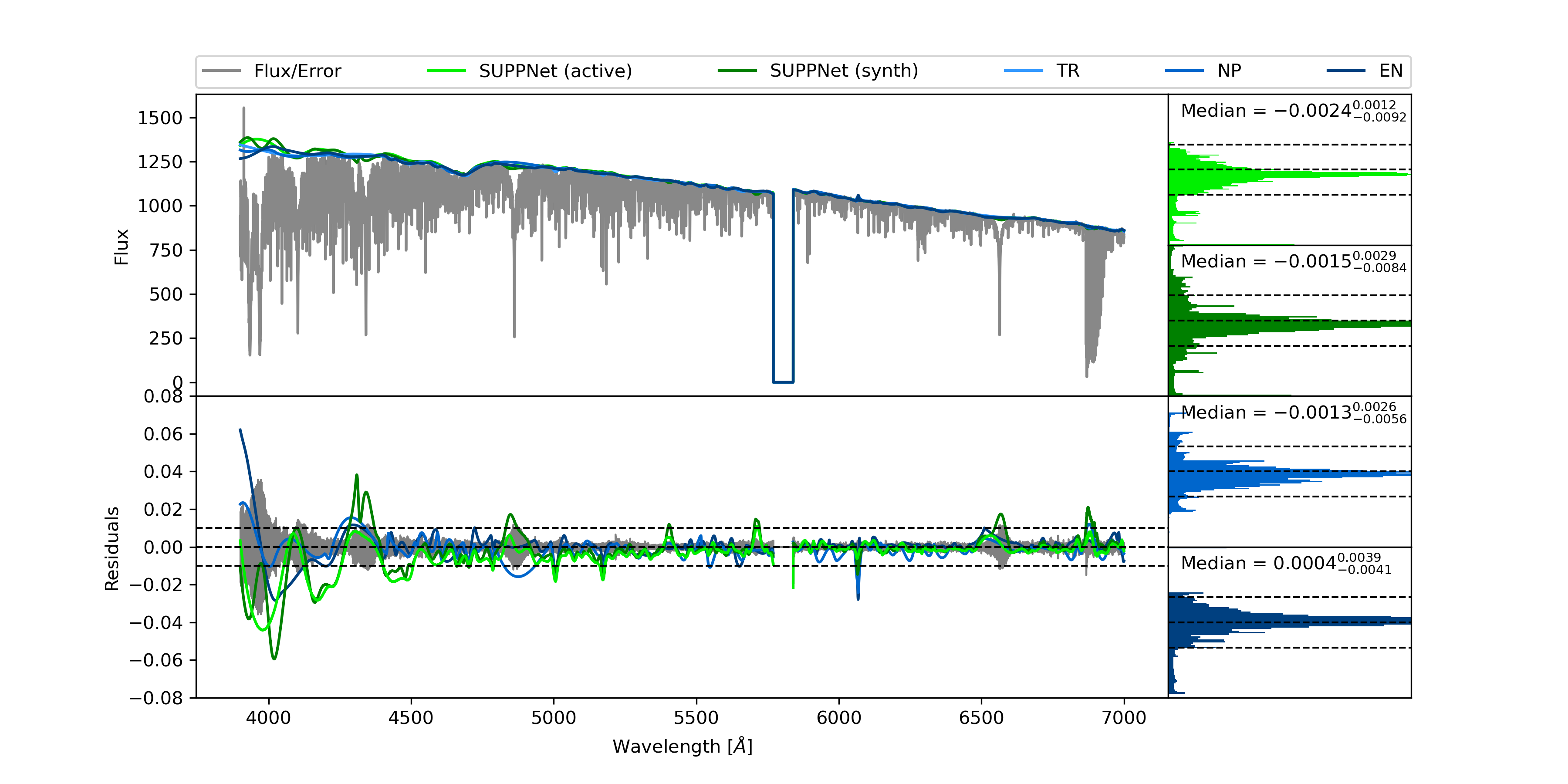}}
\caption{Comparison of normalisation quality on the example of \textbf{HD\,37495 (F4\,V)} star with two versions of the proposed method (SUPPNet active and synth) and manual normalisation done independently by three different people (TR, NP, and EN). The left upper panel shows original flux with all fitted \textit{pseudo-continua}. The left lower panel shows residuals of normalised fluxes relative to TR normalisation. The right panel presents histograms of all mentioned residuals with median, 15.87, and 84.13 percentiles.}
\label{fig:37495_manual}
\end{figure*}

\begin{figure*}
\resizebox{\hsize}{!}
{\includegraphics[clip]{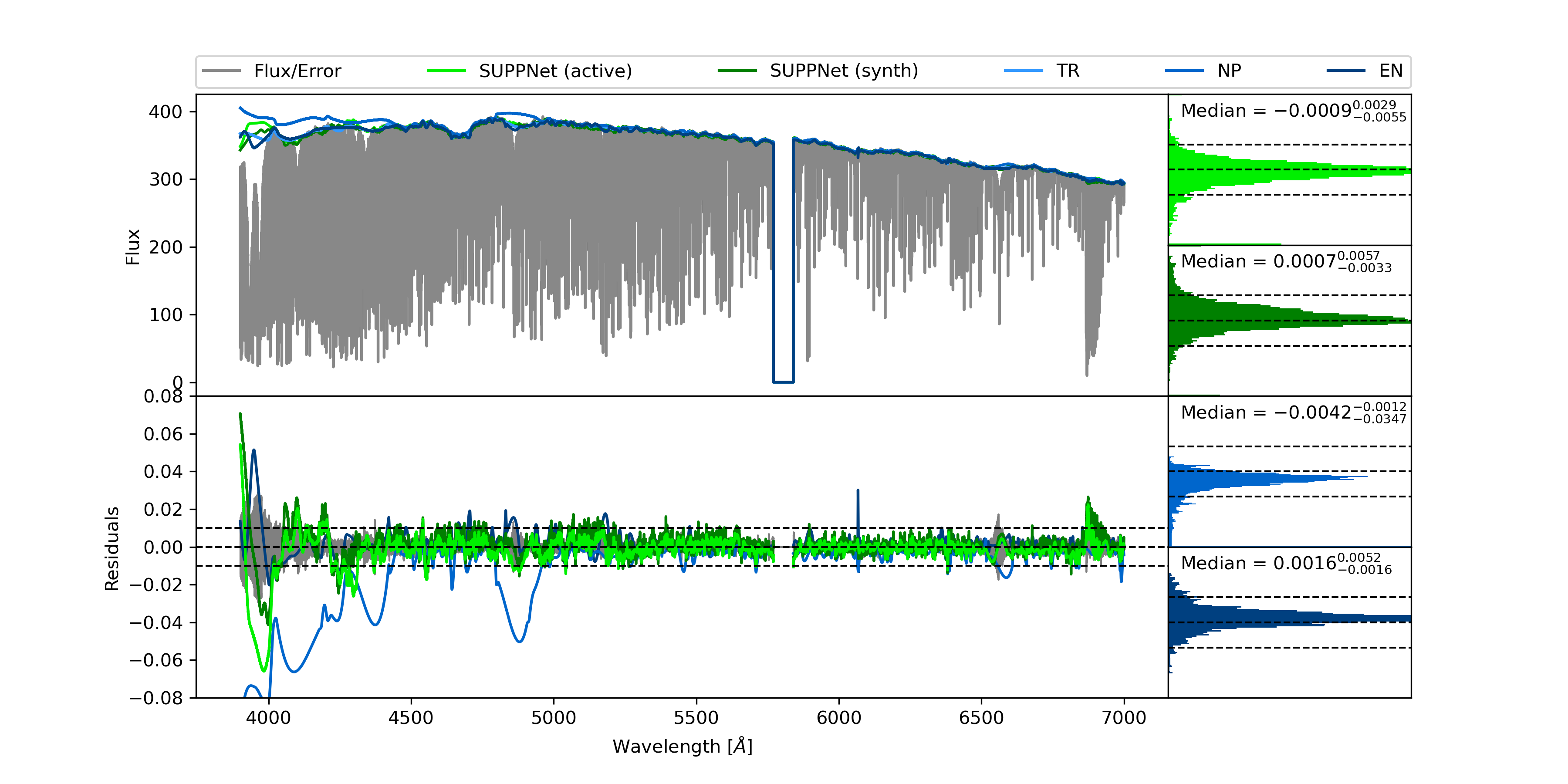}}
\caption{Comparison of normalisation quality on the example of \textbf{HD\,59967 (G4\,V)} star with two versions of the proposed method (SUPPNet active and synth) and manual normalisation done independently by three different people (TR, NP, and EN). The left upper panel shows original flux with all fitted \textit{pseudo-continua}. The left lower panel shows residuals of normalised fluxes relative to TR normalisation. The right panel presents histograms of all mentioned residuals with median, 15.87, and 84.13 percentiles.}
\label{fig:59967_manual}
\end{figure*}

\begin{figure*}
\resizebox{\hsize}{!}
{\includegraphics[clip]{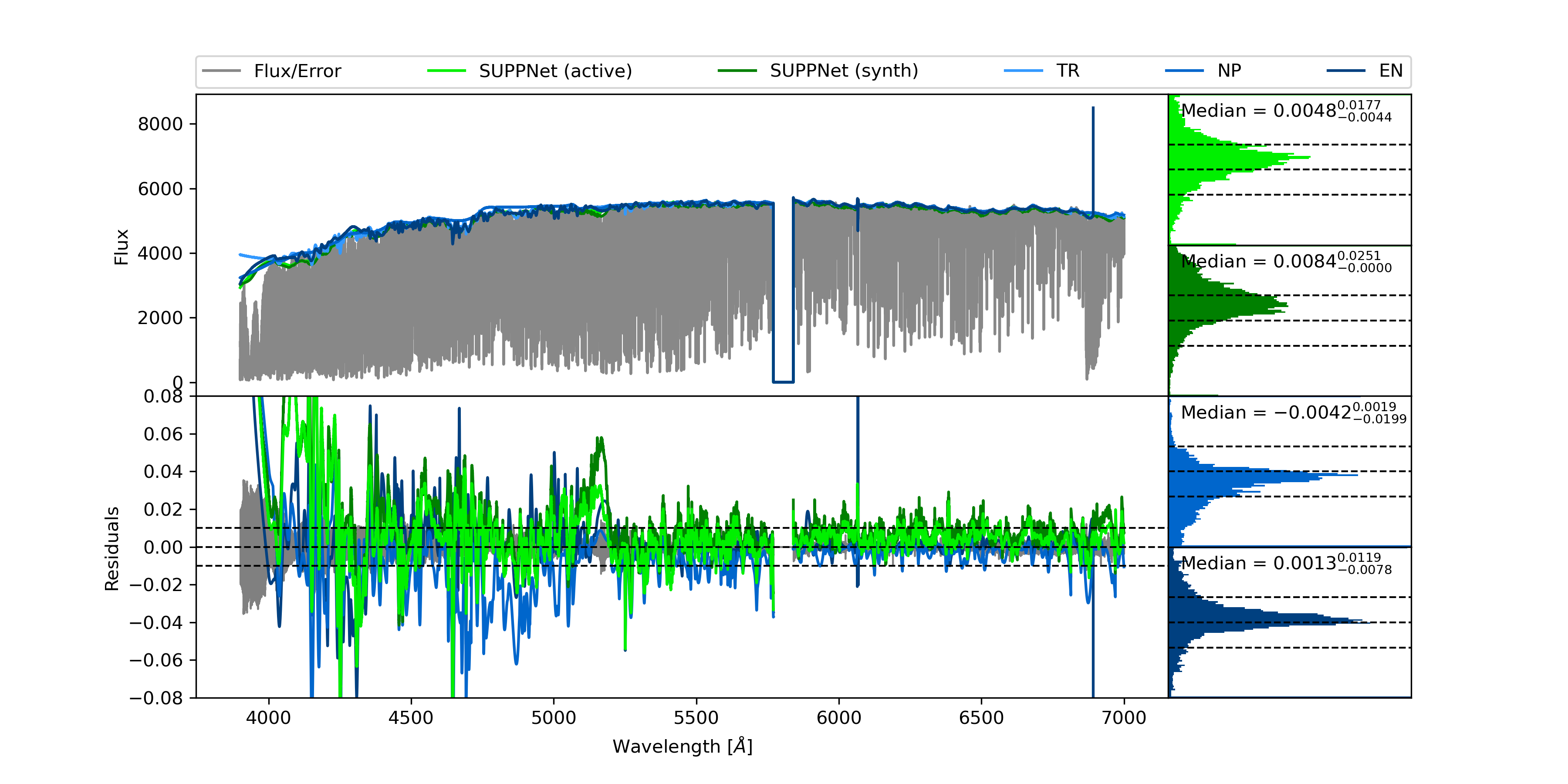}}
\caption{Comparison of normalisation quality on the example of \textbf{HD\,25069 (K0\,III)} star with two versions of the proposed method (SUPPNet active and synth) and manual normalisation done independently by three different people (TR, NP, and EN). The left upper panel shows original flux with all fitted \textit{pseudo-continua}. The left lower panel shows residuals of normalised fluxes relative to TR normalisation. The right panel presents histograms of all mentioned residuals with median, 15.87, and 84.13 percentiles.}
\label{fig:25069_manual}
\end{figure*}

\section{Resolution, rotational velocity, and noise}
\label{appendix:noise}

\begin{figure}
\resizebox{\hsize}{!}
{\includegraphics[clip]{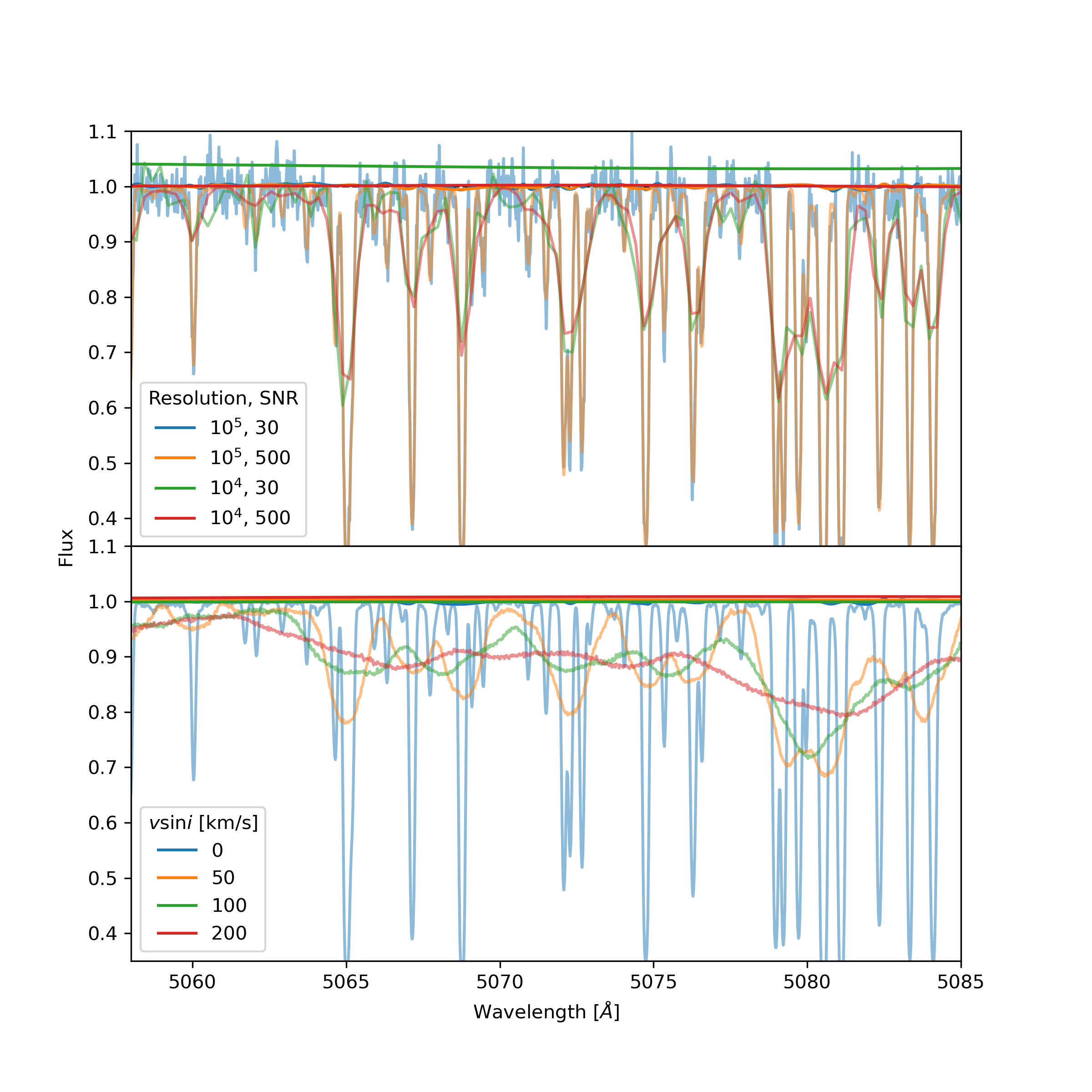}}
\caption{Normalised spectrum and \textit{pseudo-continuum} predicted by SUPPNet (active) with default sampling and smoothing parameters. As flux is already normalised, \textit{pseudo-continuum} should equal 1 in the whole domain. \textbf{The upper panel} shows the influence of noise and resolution on normalisation results. The results are generally consistent with the exception of noisy, $\textrm{SNR}=30$, medium resolution, $R=10^4$, spectra (see the green lines). In such a case the \textit{pseudo-continuum} is placed too high. It is worth mentioning that such low resolution is outside of the training data domain where the resolution was not lower than $4\times10^4$. This limitation can be overcome by increasing a sampling step or by extending the training set to include lower resolution spectra. \textbf{The lower panel} shows the influence of the projected rotational velocity on the normalisation result. The predictions are generally consistent and differ by less than 0.01.}
\label{fig:resolution_and_vsini_influence}
\end{figure}

The results of resolution, rotation velocity, and noise influence on predicted \textit{pseudo-continuum} are summarised in Fig.\,\ref{fig:resolution_and_vsini_influence}

\section{Codes}
\label{appendix:codes}

\subsection{SUPPNet -- online}

An online version of the SUPPNet method is available: \url{https://git.io/JqJhf}. It is a fully front-end JavaScript application that gives access to the basic features of SUPPNet. It is a simple way to experiment with the proposed method and is intended to be the first choice for users who do not need all features of the full version of the code and for whom the performance offered by the website is sufficient. Additionally, the webpage contains interactive versions of plots gathered in this work.

Its basic limitations are a predetermined value of the sliding window shifts (equals 1024 samples), and a missing feature of adaptive spline smoothing which is replaced with Savicky-Golay filtering \citep{1964AnaCh..36.1627S} in the online version. Users who are interested in automatised normalisation of big sets of spectra are encouraged to use the Python version of the code.

\subsection{SUPPNet -- Python version}

The Python version of SUPPNet includes all features for spectrum normalisation, with relatively easy access to GPU acceleration for users who need top performance. It is available on GitHub \url{https://github.com/RozanskiT/suppnet}.

\subsection{HANDY}

HANDY\footnote{\url{https://rozanskit.com/HANDY/}} is an interactive tool for manual spectrum normalisation. \textit{Pseudo-continuum} fitting in HANDY is based on manual selection of parts of the continuum, to which the Chebyshev polynomial of the selected order is then fitted in the defined areas. Akima spline functions are used between these areas. Additionally, HANDY includes the possibility to interpolate spectrum on the predefined grid of synthetic spectra, wraps SYNTHE/ATLAS codes \citep{kurucz1970atlas}, for spectrum synthesis that gives access to lines identification lists and includes radial velocity correction unit. Mentioned features make it a handy tool for initial spectrum exploration and/or manual normalisation.

\end{appendix}

\end{document}